\documentclass[twocolumn,showpacs,preprintnumbers,amsmath,amssymb]{revtex4}

\usepackage{graphicx}
\usepackage{dcolumn}
\usepackage{bm}

\begin{document}

\preprint{JLAB-THY-04-273}

\title{Scattering of shock waves in QCD}

\author{Ian Balitsky}
\affiliation{
Physics Dept., ODU, Norfolk VA 23529, \\
and \\
Theory Group, Jlab, 12000 Jeffeson Ave, Newport News, VA 23606
}
\email{balitsky@jlab.org}

\date{\today}

\begin{abstract}
The cross section of heavy-ion collisions is represented as a double
functional integral with the saddle point being the classical
solution of the Yang-Mills equations with boundary conditions/sources in
the form of two shock waves corresponding to the two colliding ions.
I develop the expansion of this classical solution in powers of the
commutator of the Wilson lines describing the colliding particles
and calculate the first two terms of the expansion.
\end{abstract}


\maketitle

\section{\label{sec:level1}Introduction }

Viewed from the center of mass frame,
a typical high-energy hadron-hadron scattering
looks like a collision
of two shock waves (see Fig. 1).  
\begin{figure}
\includegraphics{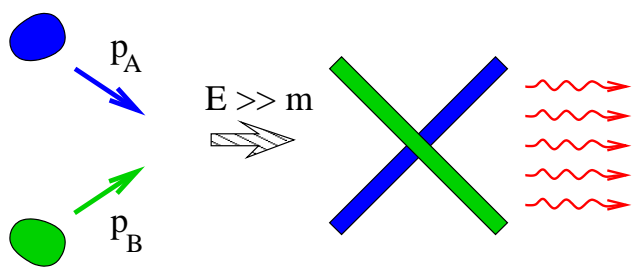}
\caption{\label{fig:1}High-energy scattering as a collision of two shock waves.}
\end{figure}
Indeed, due to the Lorentz contraction the two hadrons shrink into
thin ``pancakes'' which collide producing the final state particles.
The main question is the field/particles produced by the collision of two
shock waves. On the theoretical side, this question is related to the problem of 
high-energy effective action and to the ultimate problem of the small-$x$ physics -
unitarization of the BFKL pomeron and the Froissart bound in QCD 
(see \cite{book1,book2}). 
On more practical terms, the immediate result of the scattering of the two shock waves 
gives the initial conditions for the formation of a quark-gluon plasma studied in 
the heavy-ion collisions at RHIC (see e.g. the review \cite{mclectures}).

The collision of QCD shock waves can be treated using semiclassical methods.
 The basic idea is that at high energy the density of partons
in the transverse plane becomes sufficiently large to give the hard scale necessary for the
application of perturbation theory \cite{mvmodel,nncoll}.
The arguments in favor of this are based on the idea 
of parton saturation at high energies \cite{GLR, muchu,mu90}.
Consider a single shock wave - hadron, moving at a high speed (in  the c.m. frame).
The energetic hadron emits more and more gluons and the gluon parton density 
increases rapidly with energy. This cannot go forever - at some point the recombination of
partons balances the emission and partons reach the state of
saturation with the charactristic transverse momenta (the ``saturation scale'') being
$Q_s\sim e^{c\eta}$ where $\eta$ is the rapidity 
\cite{mu99,mabraun,iancu02,tolpa}. 
Such an energetic
shock wave with large density of color charge is called the Color Glass Condensate 
\cite{mvmodel,CGC}.

\begin{figure*}
\includegraphics{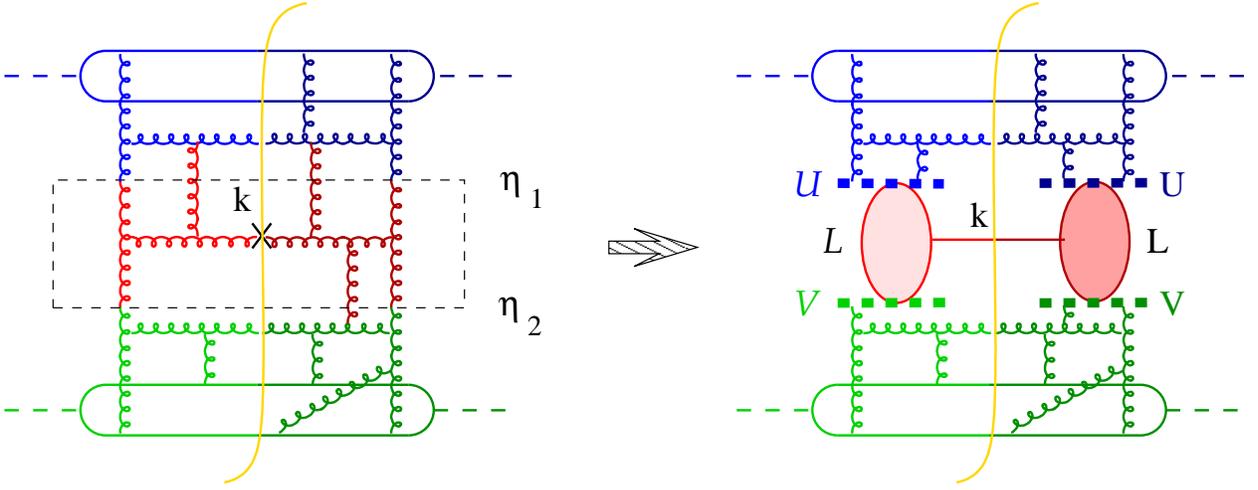}
\caption{\label{fig:2}Integration over the central rapidity gluons in a semiclassical
approximation.}
\end{figure*}
Within the semiclassical approach,
 the problem of scatering of two shock waves can be reduced to the solution of classical
 YM equations with sources being the shock waves \cite{nncoll} (see also \cite{prd99}). 
At present, these equations have not been solved. 
There are two approaches discussed in current
literature: numerical simulations \cite{krasvenu} and expansion in the strength of 
one of the shock waves \cite{kovmu99,kop,wkov}.

Note that the collision of a weak and a strong shock waves corresponds to the deep inelastic 
scattering from a
nucleus (while the scattering of two strong shock waves describes a nucleus-nucleus collision).
In the present paper I formulate the problem of scattering of shock waves, find the
boundary conditions for the double functional integral for the cross section and 
decribe the expansion in the commutators of two shock waves
equivalent to the expansion in strength of one of the waves. 
The main technical result is the calculation of the 
second term of this expansion (the first term can be restored from the current literature)

The paper is organized as follows. Sec. 2 is a more formal introduction: I outline 
the idea of the factorization of the hadron-hadron 
cross section into the formation of two shock waves and their scattering.
In Sec. 3 I discuss the rapidity  factorization and  define what is a scattering of QCD
shock waves.
In Sec. 4 I find the Lipatov vertex of the
gluon emission and in Sec. 5 reproduce the $k_T$-factorization valid in the first order 
in the commutator expansion
(for the $pA$ scattering). In Sec. 6 I obtain the first-order effective action  
and reproduce
the non-linear equation for the small-$x$ evolution of Wilson lines. Sec. 7 outlines the
calculation of the second-order classical field in the while the details of the calculation
are given in the Appendices A-C. The explicit form of the vertex of gluon emission 
by two Wilson lines in the shock-wave background is presented in the Appendix D. 

\section{Basic idea: two-step integration over rapidity}

In this section I outline how the hadron-hadron collision at high energy is related to the 
scattering of shock waves. The basic idea of the approach is the  two-step integration over 
rapidity in
the double functional integral for the cross section. At first, let us define this integral.
 
\subsection{
\label{sec:level2}Double functional integral for the cross section.}

 A total cross section is a product of an amplitude and a complex conjugate
amplitude so the functional intergral for the cross section has double set of fields: to the right of
the cut and to the left of the cut. A typical functional integral has the form:
\begin{equation}
\int\!D{\cal A} DA {\cal J}(p_A){\cal J}(p_B)e^{-iS({\cal A})}~
J(-p_A)J(-p_B)~e^{iS(A)}
\label{doubfun}
\end{equation}
where the currents $J(p_A)$ and $J(p_B)$ describe the two colliding 
particles (say, photons). Throughout the paper, fields to the left of the cut will be 
represented by the calligraphic letters while those to the right of the cut by usual letters.
The boundary conditions are such that the fields
$A$ and ${\cal A}$ coincide at $t\rightarrow\infty$, reflecting the summation
over the final states implied in the definition of a total cross section.
 The propagators for such functional integral reproduce
the Cutkovsky rules (cf. ref. \cite{keld}) :
\begin{eqnarray}
&&\hspace{-3mm}
\langle A^a_\mu(x)A^b_\nu(y)\rangle=
g_{\mu\nu}\delta^{ab}\!\int\!{d^4k\over 16\pi^4}e^{-ik(x-y)}{-i\over k^2+i\epsilon}
\label{barepropagators}\\
&&\hspace{-3mm}\langle{\cal A}^a_\mu(x){\cal A}^b_\nu(y)\rangle
=g_{\mu\nu}\delta^{ab}\!\int\!{d^4k\over 16\pi^4}e^{-ik(x-y)}
{i\over k^2-i\epsilon}
\nonumber\\
&&\hspace{-3mm}\
\langle{\cal A}^a_\mu(x)A^b_\nu(y)\rangle
=-g_{\mu\nu}\delta^{ab}\!\int\!{d^4k\over 16\pi^4}e^{-ik(x-y)}
2\pi \delta(k^2)\theta(k_0)
\nonumber
\end{eqnarray}

 We are interested in the number of gluons produced
per unit rapidity which is given by the average of the 
creation operator over the final state 
$\langle a(k_\perp,\eta)a^\dagger(k_\perp,\eta)\rangle$ (see the discussion 
in ref. \cite{jkmh}). In terms of functional
integrals this can be rewritten as
\begin{eqnarray}
&&\hspace{0mm}
n_g(k_\perp,\eta)~=~\lim_{k^2\rightarrow 0}\!\int\!D{\cal A} 
DA {\cal J}(p_A){\cal J}(p_B)
\label{dubfun}\\
&&\hspace{0mm}\times~e^{-iS({\cal A})}~
k^2{\cal A}^a_i(k)k^2A^a_i(-k)e^{iS(A)}J(-p_A)J(-p_B)
\nonumber
\end{eqnarray}
Throughout the paper, the sum over the Latin indices $i,j...$ runs over the two 
transverse components (while the sum over Greek indices runs over the 
four components as usual). 
As we shall see below, the Lipatov vertex of gluon emission 
$R_\mu(k)=\lim_{k^2\rightarrow 0} A_\mu(k)$ is transverse: $k_\mu R^\mu(k)=)$ so
we can replace $A_i{\cal A}_i$ in Eq. (\ref{dubfuna}) by the sum 
over all four indices
\begin{eqnarray}
&&\hspace{0mm}
n_g(k_\perp,\eta)~=~-\lim_{k^2\rightarrow 0}\!\int\!D{\cal A} 
DA {\cal J}(p_A){\cal J}(p_B)
\label{dubfuna}\\
&&\hspace{0mm}\times~e^{-iS({\cal A})}~
k^2{\cal A}^a_\mu(k)k^2A^{\mu}(-k)e^{iS(A)}J(-p_A)J(-p_B)
\nonumber
\end{eqnarray}

\subsection{
\label{sec:2step}Two-step integration.}

The integration over the gluon fields in the functional integral (\ref{dubfuna}) will be
done in two steps according to the rapidity of the gluons. 
Let us introduce two rapidities
$\eta_1$ and $\eta_2$ such that $\eta_A>\eta_1>\eta_2>\eta_B$.
Consider a typical Feynman diagram for the gluon production (\ref{dubfuna})
shown in Fig. \ref{fig:2}. At first, we integrate over the fields in 
the central range of rapidity $\eta_1>\eta>\eta_2$ and leave the fields with $\eta>\eta_1$ and 
$\eta<\eta_2$ in the form of external shock waves. 
In the semiclassical approximation there is only one gluon emission 
described by the Lipatov vertex - Fourirer transform of the classical field at the mass
shell. The result of the integration is the product of 
two Lipatov vertices which depend on that ``external'' fields. It is easy to see
that Lipatov vertices depend on these gluon fields through Wilson lines 
- infinite gauge links ordered along the straight lines collinear to $\eta_1$ and $\eta_2$.
Indeed, in the target frame gluons with rapidities $\eta_1>\eta>\eta_2$ are very
fast so their propagators in the background of ``target'' gluons reduce to
the gauge factor ordered along the straght line classical trajectory. 

Thus, in the semiclassical approximation we get (see Fig. \ref{fig:2})
\begin{eqnarray}
&&\hspace{-7mm}
n_g(k_\perp;\eta)~=~\int\! D{\cal A}DA D{\cal B}DB{\cal J}_AJ_Ae^{-iS({\cal A})+iS(A)}  
\label{enge}\\
&&\hspace{-7mm}  {\cal J}_BJ_Be^{-iS({\cal B})+iS(B)}
R(k;V,U,{\cal V},{\cal U})e^{i\int d^2z (V_iU^i-{\cal V}_i{\cal U}^i)}
\nonumber
\end{eqnarray}
where $R(k;V,U,{\cal V},{\cal U})$ is a product of two Lipatov vertices
\begin{eqnarray}
&&\hspace{-7mm}
R(k;V,U,{\cal V},{\cal U})~=~-R_\mu(k;V,U){\cal R}^\mu(-k,{\cal V},{\cal U})
\label{er}
\end{eqnarray}
The Lipatov vertex $R_\mu(k;V,U)=\lim_{k^2\rightarrow 0}k^2A_\mu(k;U,V)$ 
is an amplitude of the emission of a gluon 
with momentum $k$ by the two Wilson lines $U$ and $V$. It depends on the
gauge, but the product of two Lipatov vertices (\ref{er}) is gauge-invariant
due to the property $k_\mu R^\mu(k)=0$. 
Here  ${\cal A},A$ are fields with rapidities $\eta>\eta_1$ and 
${\cal B},B$ with rapidities 
$\eta<\eta_2$. The Wilson lines  $V$ and $U$ are made from $A$ and $B$
fields, respectively: 
\begin{eqnarray}
&&\hspace{-2mm}
V(x_\perp)=[\infty n_1+x_\perp,-\infty n_1+x_\perp],
\nonumber\\
&&\hspace{-2mm} 
U=[\infty n_2+x_\perp,-\infty n_2+x_\perp]
\label{usvis}
\end{eqnarray}
where $n_1$ and $n_2$ are the unit vectors corresponding to rapidities $\eta_1$ and
$\eta_2$ while $[x,y]$ is a shorthand notation for the straight-line ordered 
gauge linkconnecting points $x$ and $y$: 
\begin{equation}
[x,y]\equiv Pe^{ig\!\int_0^1\! du (x-y)^\mu A_\mu(ux +(1-u)y)}
\label{glink}
\end{equation}

It may seem that the result (\ref{enge})  depends on the artificial ``rapidity divides''
$\eta_1$ and $\eta_2$. 
This dependence should be canceled by the 
gluon ladder on the top of the Lipatov vertices which is outside the 
semiclassical approximation. 
The common belief is that the all the evolution can be 
attributed to either upper or lower sector: 
to calculate $n_g(k_\perp,\eta)$, we choose $\eta_1$ and $\eta_2$ such that both
$\alpha_s(\eta_1-\eta),\alpha_s(\eta-\eta_2)\ll 1$ so the part of the evolution
between $\eta_1$ and $\eta_2$ can be neglected and  we can use the semiclassical aproximation
for the functional integral (\ref{enge}) over the region of rapidity 
$\eta_1>\eta>\eta_2$
(see e.g. the review \cite{mclectures}). Eventually, after solving the classical 
Yang-Mills equations for this functional integral,
 we can put $\eta_1=\eta$ in $V$ and 
$\eta_2=\eta$ in U.   
The independent evolution in the upper (or lower) sectors leads to the parton saturation
so the shock waves $U$ and $V$ have a form of Color Glass Condensate \cite{CGC}. 

To find the Lipatov
vertices one needs to solve the classical YM equation with the sources proportional to
shock waves $U$ and $V$. 
As we mentioned, these equations have not been solved yet and there are two approaches 
discussed in the literature: numerical simulations and expansion in the strength of one of
the shock waves. In this paper I develop the second approach in a ``symmetric'' way
as an expansion in commutators $[U,V]$, and calculate the second term of the expansion
(the first one can be restored from the literature).

\section{Rapidity factorization and scattering of the shock waves}

\subsection{Rapidity factorization}

In this section we define the scattering of the shock waves 
using the rapidity factorization
developed in \cite{prl,prd99}.
Consider a functional integral for the cross section (\ref{doubfun})
and take some ``rapidity
divide'' $\eta_1$ such that $\eta_A>\eta_1>\eta_B$.

Throughout the paper, we use Sudakov variables
\begin{equation}
k~=~\alpha p_1+\beta p_2+k_\perp
\label{sudakov}
\end{equation}
and the notations 
\begin{eqnarray}
&&\hspace{0mm}
x_\bullet=p_1^\mu x_\mu=\sqrt{s\over 2}x^-,~~~~x^-={1\over\sqrt{2}}(x^0-x^3)
\nonumber\\
&&\hspace{0mm}
x_\ast=p_2^\mu x_\mu=\sqrt{s\over 2}x^+,~~~~x^+={1\over\sqrt{2}}(x^0+x^3)
\label{astbullet}
\end{eqnarray}
Here $p_1$ and $p_2$ are the light-like vectors close to $p_A$ and $p_B$:
$p_A=p_1+{p_A^2\over s}p_2$, $p_B=p_2+{p_B^2\over s}p_1$. 

Let us integrate first over the fields with the rapidity $\eta>\eta_1$. From the viewpont 
of such particles, the gluons with $\eta<\eta_1$ shrink to a shock wave so 
the result of the integration is presented by Feynman diagrams in the
shock-wave background. In the covariant gauge, the shock-wave has the
only non-vanishing component $A_\bullet$ which is concentrated near $x_\ast=0$.
A typical Green function $G(x,y)$ at $x_\ast,y_\ast<0$ in the background-Feynman gauge
has the form \cite{mobzor}
\begin{eqnarray}
&&\hspace{-2mm}
\langle {\cal A}(x)A(y)\rangle ~=
\nonumber\\
&&\hspace{-2mm}~\int dz\delta({2\over s}z_\ast)(x|{-1\over p^2-i\epsilon }
\Big\{2\alpha g_{\mu\nu}{\cal U}_zU_z+
{4i\over s}(\partial_\mu({\cal U}_zU_z)p_{2\nu}
\nonumber\\
&&\hspace{-2mm}~-\mu\leftrightarrow\nu)-~{4p_{2\mu}p_{2\nu}\over \alpha s^2}
\partial_\perp^2({\cal U} U)_z\Big\}
(z|{1\over p^2+i\epsilon }|y)
\label{greenfun}
\end{eqnarray}
where ${\cal U}_z=[\infty p_1+z_\perp,-\infty p_1+z_\perp]$ is made form
the left fields ${\cal A}$ while $U_z=[\infty p_1+z_\perp,-\infty p_1+z_\perp]$
is made from $A$'s.

Similarly to the case of the usual functional integral for the amplitude, 
in order to write down factorization we need to rewrite the shock wave in 
the temporal gauge $A_0=0$. In such gauge the shock-wave background 
has the form 
\begin{eqnarray}
&&\hspace{0mm}
{\cal A}_i={\cal U}_i\theta(-x_\ast),~~{\cal A}_\bullet={\cal A}_\ast=0,
\nonumber\\
&&\hspace{0mm}
A_i=U_i\theta(-x_\ast),~~A_\bullet=A_\ast=0
\label{shok}
\end{eqnarray}
where 
\begin{equation}
{\cal U}_i\equiv {\cal U}{i\over g}\partial_i{\cal U}^\dagger,~~~~~
U_i\equiv U{i\over g}\partial_iU^\dagger
\label{defui}
\end{equation}
are the pure gauge fields (filling the
half-space $x_\ast<0$).
 Note that the choice (\ref{shok}) is different from
the choice $A_i=U_i\theta(x_\ast)$ adopted in \cite{prd99,mobzor}. The reason is the
following: when we calculate the amplitude, it is natural to use the redundant 
gauge rotation
to get rid of the field at $t=-\infty$ 
(althougth the choice $t=\infty$ is equally possible).
On the contrary, for a double functional integral of the form (\ref{doubfun}), 
we have two independent integrations at $t=-\infty$ and it is 
impossible to eliminate both of them. We can however gauge away 
fields at $t=\infty$ because of the boundary condition $A={\cal A}|_{t=\infty}$. 
(Stricltly speaking, we cannot gauge away all the fields;
 what we can do is to forbid a pure gauge fields $t=\infty$ which 
 is enough for our purposes
 since it puts forward the choice (\ref{shok}) over the 
 choice  $U_i\theta(x_\ast)$).

The Green functions in the background (\ref{shok}) differ from those 
of (\ref{greenfun}) 
by a simple gauge rotation. Their explicit form is 
presented in the Appendix \ref{sect:props}. 

The generating functional for the Green functions in the Eq. (\ref{shok}) 
background is obtained by the generalization of the generating functional
of \cite{mobzor} to the case of a double functional integral: 
\begin{eqnarray}
&&\hspace{0mm}
\int\!D{\cal A} DA {\cal J}(p_A)J(-p_A)e^{-iS({\cal A})
+i\!\int\! d^2z_\perp 
 (\infty,{\cal F}_{\ast i},\infty)^a_z{\cal U}^{ai}_z}
\nonumber\\
&&\hspace{0mm}~e^{iS(A)
-i\!\int\! d^2z_\perp 
 (\infty,F_{\ast i},\infty)^a_zU^{ai}_z}
 \label{generfun}
\end{eqnarray}
where the Wilson-line operator
\begin{eqnarray}
&&\hspace{-3mm}
(\infty,{\cal F}_{\mu i},\infty)^a\equiv 
\\
&&\hspace{-3mm}
2{\rm tr}~t^a\!\int_{-\infty}^\infty\! \!du~
[\infty p_2,up_2]_z{\cal F}_{\mu i}(up_2+z_\perp)[up_2,-\infty p_2]_z,
\nonumber
\end{eqnarray}
is made from the ``left'' fields ${\cal A}_\mu$ and the operator
\begin{eqnarray}
&&\hspace{-3mm}
(\infty,F_{\mu i},\infty)^a\equiv
\nonumber\\
&&\hspace{-3mm} 
2{\rm tr}~t^a\!\int_{-\infty}^\infty \!\!du~
[\infty p_2,up_2]_zF_{\mu i}(up_2+z_\perp)[up_2,-\infty p_2]_z
\nonumber
\end{eqnarray}
from the ``right'' fields $A_\mu$.

It is easy to see that the functional integral (\ref{generfun})
generates Green functions in the Eq. (\ref{shok}) 
background. Indeed, let us choose the gauge $A_\ast={\cal A}_\ast=0$ for
simplicity. In this gauge  
$(\infty,{\cal F}_{\ast i},\infty)^a={\cal A}_i(\infty p_2+z_\perp)-
{\cal A}_i(-\infty p_2+z_\perp)$ and 
$(\infty,F_{\ast i},\infty)^a=A_i(\infty p_2+z_\perp)-A_i(-\infty p_2+z_\perp)$.
so the functional integral (\ref{generfun})  takes the form 
\begin{eqnarray}
&&\hspace{0mm}
\int\!D{\cal A} DA {\cal J}(p_A)J(-p_A)
\\
&&\hspace{0mm}
\times~e^{-iS({\cal A})
+i\!\int\! d^2z_\perp 
({\cal A}_i(\infty p_2+z_\perp)-
{\cal A}_i(-\infty p_2+z_\perp))^a{\cal U}^{ai}}
\nonumber\\
&&\hspace{0mm}
\times~e^{iS(A)
-i\!\int\! d^2z_\perp 
 (A_i(\infty p_2+z_\perp)-A_i(-\infty p_2+z_\perp))^aU^{ai}}   
\nonumber
\end{eqnarray}
Let us now shift the fields $A_i\rightarrow A_i+\bar{A}_i$ and 
${\cal A}_i\rightarrow {\cal A}_i+ \bar{{\cal A}}_i$ where
 $\bar{A}_i=U_i\theta(-x_\ast)$ and
$\bar{\cal A}_i={\cal U}_i\theta(-x_\ast)$. 
The only non-zero components of the classical field strength in our case are 
$F_{\bullet i}=-U_i\delta({2\over s}x_\ast)$ 
so we get (for the ``right'' sector)
\begin{eqnarray}
&&\hspace{-5mm}S(A+\bar{A})~=~{2\over s}\!\int d^4zD^i\bar{F}_{i\bullet}A^\ast
-{2\over s}\int\! d^2z_\perp dz_\bullet
\nonumber\\
&&\hspace{-5mm}\times~\left.
A^i\bar{F}_{\bullet i}\right|^{x_\ast=\infty}_{x_\ast=-\infty}
+{1\over 2} A^\mu(\bar{D}^2g_{\mu\nu}-2i\bar{\cal F}_{\mu\nu})A^\nu +...
\label{afteshift}
\end{eqnarray}
 In the $A_\ast=0$ gauge the first term in the r.h.s. of
Eq. (\ref{afteshift}) vanishes while the second term cancels with the corresponding
contribution $\sim - (A_i(\infty p_2+z_\perp)-A_i(-\infty p_2+z_\perp))^aU^{ai}$ 
coming from the source in Eq. (\ref{generfun}). Similar cancellation occurs in 
the left sector so we get
\begin{eqnarray}
&&\hspace{0mm}
\int\!D{\cal A} DA {\cal J}(p_A)J(-p_A)
\label{generfuna}\\
&&\hspace{0mm}\times~e^{-iS({\cal A})
+i\!\int\! d^2z_\perp 
[{\cal A}_i(\infty p_2+z_\perp)-
{\cal A}_i(-\infty p_2+z_\perp)]^a{\cal U}^{ai}_z}
 \nonumber\\
&&\hspace{0mm}\times~e^{iS(A)
-i\!\int\! d^2z_\perp [
 A_i(\infty p_2+z_\perp)-A_i(-\infty p_2+z_\perp)]^aU^{ai}_z}
 \nonumber\\
&&\hspace{0mm} =~\int\!D{\cal A} DA {\cal J}(p_A)J(-p_A)
 \nonumber\\
&&\hspace{0mm}\times~
e^{-{i\over 2}\!\int\! d^2z{\cal A}^\mu(\bar{\cal D}^2g_\mu\nu
-2i\bar{\cal F}_{\mu\nu}){\cal A}^\nu}
e^{{i\over 2}\!\int\! d^2zA^\mu(\bar{D}^2g_\mu\nu
-2i\bar{F}_{\mu\nu})A^\nu}
\nonumber
\end{eqnarray}
which gives the Green functions in the 
Eq. (\ref{shok}) background.

To complete the factorization formula one needs to integrate 
over the remaining fields
with rapidities $\eta<\eta_1$: 
\begin{eqnarray}
&&\hspace{-3mm}
\int\!D{\cal A} DA {\cal J}(p_A){\cal J}(p_B)e^{-iS({\cal A})}~e^{iS(A)}
J(-p_A)J(-p_B)
\label{faktor}
\nonumber\\
&&\hspace{-3mm} 
=~\int\!D{\cal A} DA {\cal J}(p_A)J(-p_A)\int\!D{\cal B} DB{\cal J}(p_B)J(-p_B)
\nonumber\\
&&\hspace{-3mm}\times~
e^{-iS({\cal A})-iS({\cal B})
+i\!\int\! d^2z_\perp 
 (\infty,{\cal F}_{\mu i},\infty)^a_z
 (\infty,{\cal G}_{\nu i},\infty)^a_z e_1^\mu e_1^\nu}
\nonumber\\
&&\hspace{-3mm} 
\times~e^{iS(A)+iS(B)
-i\!\int\! d^2z_\perp 
 (\infty,F_{\mu i},\infty)^a_z
 (\infty,G_{\nu i},\infty)^a_z e_1^\mu e_1^\nu}
\end{eqnarray}
As discussed in \cite{prl,prd99,mobzor,baba03}, the slope of Wilson lines 
is determined by the 
``rapidity divide'' vector $e_{\eta_1}=p_1+e^{-\eta_1}p_2$. (From the wiewpoint
of $A$ fields, the slope $e_1$ can be replaced by $p_2$ with power accuracy so we
recover the generating functional (\ref{generfun}) with 
$U_i=(\infty,G_{\ast i},\infty)$, ${\cal U}_i=(\infty,{\cal G}_{\ast i},\infty)$).

Applying the factorization formula (\ref{faktor}) two times one gets:
\begin{eqnarray}
&&\hspace{-3mm}
-\lim_{k^2\rightarrow 0}\!\int\!D{\cal A} DA {\cal J}(p_A){\cal J}(p_B)
\label{fak2times}\\
&&\hspace{-3mm}
\times~ e^{-iS({\cal A})}~
k^2{\cal A}^{a\mu}(k)k^2A^a_\mu(-k)e^{iS(A)}J(-p_A)J(-p_B)
\nonumber\\
&&\hspace{-3mm} =~-\int\!D{\cal A} DA {\cal J}(p_A)J(-p_A)
e^{-iS({\cal A})+iS(A)}
\nonumber\\
&&\hspace{-3mm} 
\!\int\!D{\cal B} DB{\cal J}(p_B)J(-p_B)e^{-iS({\cal B})+iS(B)}
\nonumber\\
&&\hspace{-3mm} 
\times~\int\!D{\cal C} DC \lim_{p^2\rightarrow 0}p^2{\cal C}^{a\mu}(p)p^2C^a_\mu(-p)
 \exp\Big[-iS({\cal C})
 \nonumber\\
&&\hspace{-3mm}
+~iS(C)+\!\int\! d^2z_\perp \Big\{e_1^\mu e_1^\nu\Big(
 [\infty,{\cal A}_{\mu i},\infty]^a_z
 [\infty,{\cal C}_{\nu i},\infty]^a_z 
\nonumber\\
&&\hspace{-3mm}  -~[\infty,A_{\mu i},\infty]^a_z
 [\infty,C_{\nu i},\infty]^a_z
 \Big)+e_2^\mu e_2^\nu\Big(
 (\infty,{\cal C}_{\mu i},\infty)^a_z
 \nonumber\\
&&\hspace{-3mm}\times~(\infty,{\cal B}_{\nu i},\infty)^a_z
 -(\infty,C_{\mu i},\infty)^a_z
 (\infty,B_{\nu i},\infty)^a_z\Big)\Big\}\Big]
 \nonumber
\end{eqnarray}
where the slope is $e_1=p_1+e^{-\eta_1}p_2$ for the $[...]$  Wilson lines
and $e_2=p_1+e^{-\eta_2}p_2$ for the $(...)$ ones.

The functional integral over the central range of rapidity 
$\eta_1>\eta>\eta_2$ is determined by scattering of shock two shock waves:
\begin{eqnarray}
&&\hspace{-5mm}-\!
\int\!\!D{\cal C} DC \lim_{k^2\rightarrow 0}k^2{\cal C}^{a\mu}(k)k^2C^a_\mu(-k)
 \exp\Big[iS(C)-iS({\cal C})
 \nonumber\\
&&\hspace{-3mm} 
 \!\int\! d^2z_\perp \Big\{e_1^\mu \Big(
 {\cal V}^{ai}_z
 [\infty,{\cal C}_{\mu i},\infty]^a_z  -V^{ai}_z
 [\infty,C_{\nu i},\infty]^a_z
 \Big)
 \nonumber\\
&&\hspace{-3mm} +~\Big({\cal U}^{ai}_z
 (\infty,{\cal C}_{\mu i},\infty)^a_z
-U^{ai}_z
 (\infty,C_{\mu i},\infty)^a_z
 \Big)e_2^\mu \Big\}\Big]
 \label{eqn20}
\end{eqnarray}
where ${\cal V}_i= e_1^\mu[\infty,{\cal A}_{\mu i},\infty]$, 
$V_i=e_1^\mu[\infty,A_{\mu i},\infty]$, 
${\cal U}_i= e_2^\mu(\infty,{\cal B}_{\mu i},\infty)$, and
$U_i= e_2^\mu(\infty,B_{\mu i},\infty)$ are the pure-gauge ``external'' fields
(to be integrated over later).
With a power accuracy $O(m^2/s)$, we can replace $e_1$ by $p_1$ and $e_2$ by $p_2$ :
\begin{eqnarray}
&&\hspace{-5mm}-\!
\int\!\!D{\cal C} DC \lim_{k^2\rightarrow 0}k^2{\cal C}^{a\mu}(k)k^2C^a_\mu(-k)
 \exp\Big\{iS(C)-iS({\cal C})
  \nonumber\\
&&\hspace{-5mm}+~i\!\int\! d^2z_\perp \Big(
 {\cal V}^{ai}_z
 [\infty,{\cal C}_{\bullet i},\infty]^a_z
+ (\infty,{\cal C}_{\ast i},\infty)^a_z
 {\cal U}^{ai}_z \nonumber\\
&&\hspace{5mm}-~
 V^{ai}_z
 [\infty,C_{\bullet i},\infty]^a_z- 
 (\infty,C_{\ast i},\infty)^a_z
 U^{ai}_z\Big)\Big\}
 \label{integc}
\end{eqnarray}
The saddle point of the functional integral (\ref{integc}) is determined by the 
classical equations
\begin{eqnarray}
&&\hspace{-0mm}
{\delta\over\delta C_\mu^a}\Big\{S(C) 
-i\!\int\! d^2z_\perp 
 \Big(V^{ai}_z
 [\infty,C_{\bullet i},\infty]^a_z
\nonumber\\
&&\hspace{20mm} + ~(\infty,C_{\ast i},\infty)^a_z  U^{ai}_z\Big)\Big\}~=~0
 \nonumber\\
&&\hspace{-0mm}
{\delta\over\delta {\cal C}_\mu^a}\Big\{S({\cal C}) 
-i\!\int\! d^2z_\perp  \Big(
 {\cal V}^{ai}_z
 [\infty,{\cal C}_{\bullet i},\infty]^a_z
\nonumber\\
&&\hspace{20mm} + ~
(\infty,{\cal C}_{\ast i},\infty)^a_z  {\cal U}^{ai}_z\Big)~=~0
 \nonumber\\
\label{cliqs}
\end{eqnarray}
At present it is not known how to solve this equations 
(for the numerical solution see \cite{krasvenu}). In the next section 
we will develop a ``perturbation theory'' in powers of the parameter 
$[U,V]\sim g^2[U_i,V_j]\sim g^2[{\cal U}_i,{\cal V}_j]$. Note that the 
conventional perturbation theory 
corresponds to the case when $U_i,V_i\sim 1$ while the semiclassical QCD is relevant
when the fields are large ($U_i$ and/or $V_i\sim{1\over g}$).  

\subsection{Expansion in commutators of Wilson lines}

\begin{figure}
\includegraphics{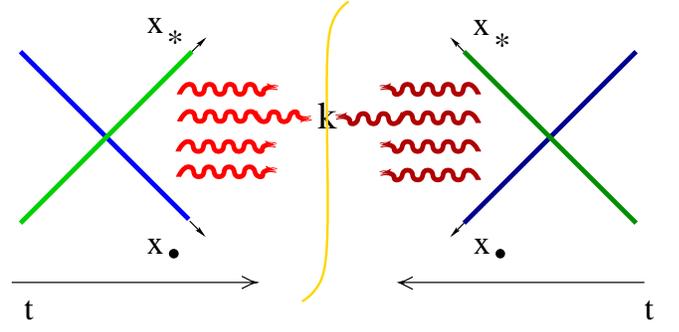}
\caption{\label{fig:3alt}Particle production
due to the scattering of the two shock waves.}
\end{figure}

The particle production due to scattering of the
two shock waves in QCD (see Fig. \ref{fig:3alt})
is determined by the functional integral (\ref{integc}) (hereafter we switch
back to the usual notation $A_\mu$ for the integration variable and $F_{\mu\nu}$ for
the field strength)
\begin{eqnarray}
&&\hspace{-4mm} 
R(p;U,V,{\cal U},{\cal V})~=~-\lim_{k^2\rightarrow 0}
\int\!\!D{\cal A} DA~k^2{\cal A}^{a\mu}(k)
\nonumber\\
&&\hspace{-4mm}
 \times~k^2A^a_\mu(-k)~\exp\Big\{-iS({\cal A})
 +i\!\int\! d^2z_\perp \Big({\cal V}^{ai}_z
 [\infty,{\cal F}_{\ast i},\infty]^a_z
  \nonumber\\
&&\hspace{-4mm}
+~(\infty,{\cal F}_{\ast i},\infty)^a_z
 {\cal U}^{ai}_z\Big)
+iS(A)-i\!\int\! d^2z_\perp \Big(
 V^{ai}_z \nonumber\\
&&\hspace{13mm}\times~
 [\infty,F_{\ast i},\infty]^a_z
  + 
 (\infty,F_{\ast i},\infty)^a_z
 U^{ai}_z\Big\}
 \label{mastegral}
\end{eqnarray}
Taken separately, the source $U_i$ creates
a shock wave $U_i\theta(-x_\ast)$ and 
$V_i$ creates $V_i\theta(-x_\bullet)$ 
(to the left of the cut, ${\cal U}_i$ generates 
the classical field ${\cal U}_i\theta(-x_\ast)$ and 
${\cal V}_i$ generates ${\cal V}_i\theta(-x_\bullet)$ ). 
In QED, the two sources $U_i$ and $V_i$ do not interact so the sum of 
the two shock waves
\begin{eqnarray}
&&\hspace{-5mm}
\bar{\cal A}_i^{(0)}={\cal U}_i\theta(-x_\ast)+{\cal V}_i\theta(-x_\bullet),~~~
\bar{\cal A}_\bullet^{(0)}=\bar{\cal A}_\ast^{(0)}=0
\nonumber\\
&&\hspace{-5mm}\bar{A}_i^{(0)}=U_i\theta(-x_\ast)+V_i\theta(-x_\bullet),~~~
\bar{A}_\bullet^{(0)}=\bar{A}_\ast^{(0)}=0
 \label{shoksum}
\end{eqnarray}
is a classical solution to the set of equations (\ref{cliqs}). In QCD, 
the
interaction between these two sources is described by the commutator $g[U_i,V_k]$ 
(the coupling constant $g$  corresponds to the three-gluon vertex).  
It is natural to take the trial
configuration in the form of a sum of the two shock waves
and expand the ``deviation'' of the full QCD solution from
the QED-type ansatz (\ref{shoksum}) in powers of commutators $[U,V]$. 
To carry this out, one
 shifts $A\rightarrow A+\bar{A}_i^{(0)}$, 
${\cal A}\rightarrow {\cal A}+\bar{\cal A}_i^{(0)}$ in the functional integral 
(\ref{mastegral})
and obtains
\begin{eqnarray}
&&\hspace{0mm} 
R(k;U,V,{\cal U},{\cal V})~=~-\int\!D{\cal A} DA~
\lim_{k^2\rightarrow 0} k^2{\cal A}^{a\mu}(k)
\nonumber\\
&&\hspace{0mm}
\times~k^2A^a_\mu(-k)
 \exp\Big\{i\!\int\! d^4z\Big(-
({1\over 2}{\cal A}^\mu\bar{\cal D}_{\mu\nu}{\cal A}^\nu 
  \nonumber\\
&&\hspace{0mm}+~
 {1\over 2} A^\mu \bar{D}_{\mu\nu}A^\nu
 -{\cal T}^\mu{\cal A}_\mu+T^\mu A_\mu\Big)\Big\}~,
 \label{mastegral1}
\end{eqnarray}
Here $D_{\mu\nu}=D^2(\bar{A})g_{\mu\nu}-2i \bar{F}_{\mu\nu}$
(${\cal D}_{\mu\nu}=D^2(\bar{\cal A})g_{\mu\nu}-2i\bar{\cal F}_{\mu\nu}$) 
is the inverse propagator in the background-Feynman gauge
\footnote{Strictly speaking, one should add to the operator $D_{\mu\nu}$ 
the second variational derivative of the source (\ref{seconder}) 
which contributes to the
$\partial_\perp^2{\cal U}^\dagger U$ term in the Green function, 
see ref.\cite{npb96}}
and
$T_\mu({\cal T}_\mu$ is the linear term for the trial configuration
(\ref{shoksum})
\begin{eqnarray}
&&\hspace{0mm}T_\mu~=~
-D^k F_{\mu k}\theta(-z_\ast)\theta(-z_\bullet)
\label{Ts}\\
&&\hspace{0mm} 
+~ig[U_i,V^i]\Big(p_{1\mu}\theta(z_\ast)\delta(z_\bullet)
-p_{2\mu}\theta(z_\bullet)\delta(z_\ast)\Big)
\nonumber\\
&&\hspace{0mm} 
 {\cal T}_\mu~=~-
D^k{\cal F}_{\mu k}\theta(-z_\ast)\theta(-z_\bullet)
\nonumber\\
&&\hspace{0mm} 
+~ig[{\cal U}_i,{\cal V}^i]\Big(p_{1\mu}\theta(z_\ast)\delta(z_\bullet)-
p_{2\mu}\theta(z_\bullet)\delta(z_\ast)\Big)
\nonumber 
\end{eqnarray}
where $\bar{F}_{ik}=-ig[U_i,V_k]-i\leftrightarrow k$ 
(and $\bar{\cal F}_{ik}=-ig[{\cal U}_i,{\cal V}_k]-i\leftrightarrow k$).

Expanding in powers of $T$(${\cal T}$) in the functional integral (\ref{mastegral1}) 
one gets the set of Feynman diagrams in the external fields (\ref{shoksum}) with the sources
(\ref{Ts}). The parameter of the expansion is $g^2[U_i,V_j]$ ($\sim [U,V]$, see Eq.
(\ref{defui})).

\section{Classical fields and Lipatov vertex in the first order in $[U,V]$}

The general formulas for the classical solution in the first order in 
$[U,V]$ ($[{\cal U},{\cal V}]$) have the form
\begin{eqnarray}
\hspace{0mm}
\bar{A}^{(1)a}_\mu(x) &=&
i\!\int\! d^4z
\Big(\langle A^a_\mu(x)A^{b\nu}(z)\rangle_{\bar{A},\bar{\cal A}} T^b_\nu(z)
\nonumber\\&-&~~~~~
\langle A^a_\mu(x){\cal A}^{b\nu}(z)\rangle_{\bar{A},\bar{\cal A}} {\cal T}^b_\nu(z)
\Big)
\nonumber \\ 
\bar{\cal A}^{(1)a}_\mu(x)&=& 
i\!\int\! d^4z\Big(\langle {\cal A}^a_\mu(x)A^{b\nu}(z)\rangle_{\bar{A},\bar{\cal A}}
 T^b_\nu(z)
 \nonumber\\&-&~~~~~
\langle {\cal A}^a_\mu(x){\cal A}^{b\nu}(z)\rangle_{\bar{A},\bar{\cal A}} 
{\cal T}^b_\nu(z)
\Big)\label{1order}
\end{eqnarray}
The Green functions in the background of the Eq. (\ref{shoksum})
field can be approximated by cluster expansion
\begin{eqnarray}
&&\hspace{0mm}
\langle A_\mu(x)A^\nu(z)\rangle_{\bar{A},\bar{\cal A}}
\nonumber \\
&&\hspace{0mm} =~ 
\langle A_\mu(x)A^\nu(z)\rangle_{U,{\cal U}}
+\langle A_\mu(x)A^\nu(z)\rangle_{V,{\cal V}}
\nonumber \\
&&\hspace{0mm}-~
\langle A_\mu(x)A^\nu(z)\rangle_0+O([U,V],[{\cal U},{\cal V}])
\label{cluster}
\end{eqnarray}
(and similarly for other sectors) where $\langle A_\mu(x)A^\nu(z)\rangle_0$
are the perturbative propagators (\ref{barepropagators}) 
and the propagators in the background of one shock wave are given
in the Appendix \ref{sect:props}. 
As a first step we shall discuss the behavior of classical fields at
the $t=-\infty$ boundary.

\subsection{Pure gauge fields at $t=-\infty$}

Note that while each of the field $U_i$ and $V_i$ satisfies the
boundary condition $A_i(t\rightarrow -\infty)$ = pure gauge, their sum 
(\ref{shoksum}) does not. I will demonstrate now that the 
correction (\ref{1order}) restores this property so
$\bar{A}^{(0)}+\bar{A}^{(1)}$ is a pure gauge (up to $[U,V]^2$ terms). 

We need to 
prove that $\bar{D}_\mu \bar{A}^{(1)}_\nu-\mu\leftrightarrow\nu$ cancels
$\bar{F}_{ik}=-ig([U_i,V_k]-i\leftrightarrow k)$ as $t\rightarrow -\infty$ .
First, note that the contribution from terms $[U_i,V^i]$ in $T$
(and $[{\cal U}_i,{\cal V}^i]$  in ${\cal T}$) vanishes at 
$t\rightarrow -\infty$ since these
sources are located at $t\geq 0$. Also, it is easy to see that the 
Green functions $\langle{\cal A}_\mu(x)A^\nu(y)\rangle$ interpolating
between sectors also vanish at this limit (see the explicit expressions in the
Appendix \ref{sect:a1}). The only nonzero contribution to $A^{(1)}_\nu$ 
at $t= -\infty$ has the form 
\begin{eqnarray}
&&\hspace{-3mm}A^{(1)}_i(x)=g\!
\int\!d\alpha d\beta d^2z_\perp {e^{-i\alpha x_\bullet-i\beta x_\ast}\over
(\alpha-i\epsilon)(\beta-i\epsilon)}(U_xU^\dagger_z+V_xV^\dagger_z
\nonumber\\
&&\hspace{-3mm}-~1)^{ab}
(x_\perp|{p^k\over \alpha\beta s-p_\perp^2+i\epsilon}|0,z_\perp)
([U_i,V_k]_z-i\leftrightarrow k)
\nonumber \\
&&\hspace{-3mm}=~g
(x_\perp|U{p^k\over p_\perp^2}U^\dagger+V{p^k\over p_\perp^2}V^\dagger
-{p^k\over p_\perp^2}|[U_i,V_k]
-i\leftrightarrow k)
\label{ai1}
\end{eqnarray}
Throughout the paper, we use the notations
\begin{eqnarray}
&&\hspace{-1mm}(x_\perp|...|f)\equiv \int d^2z_\perp (x_\perp|...|z_\perp)f(z_\perp)
\nonumber\\
&&\hspace{-1mm}
(x|...|0,f)\equiv \int d^2z_\perp (x|...|0,z_\perp)f(z_\perp)
\end{eqnarray}
%

\begin{figure*}
\includegraphics{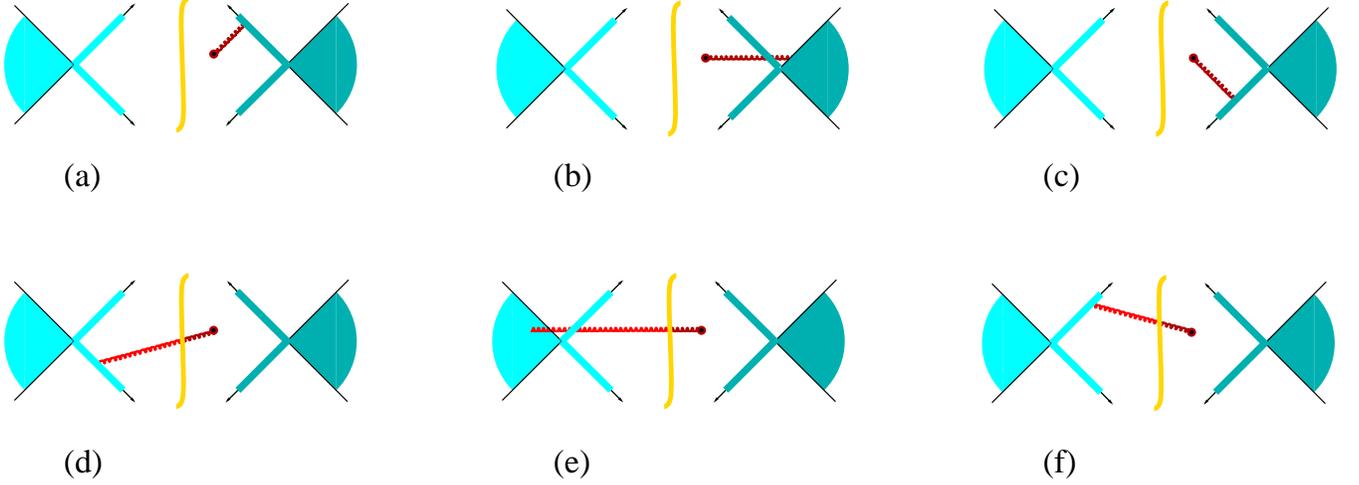}
\caption{\label{fig:5}Classical field in the first order. The shaded area represents
the linear term (\ref{Ts}).}
\end{figure*}
From Eq. (\ref{ai1}) we obtain
\begin{eqnarray}
&&\hspace{-3mm} D_i A^{a(1)}_j-i\leftrightarrow j
\\
&&\hspace{-3mm}
=~g(x_\perp|U{p_ip^k\over p_\perp^2}U^\dagger+
V{p_ip^k\over p_\perp^2}V^\dagger-{p_ip^k\over p_\perp^2}|^{ab}\bar{F}^b_{jk})
-i\leftrightarrow j
\nonumber
\end{eqnarray}
Because $\bar{F}_{jk}\sim \epsilon_{ik}$ in two dimensions 
$p_ip^k~\bar{F}_{jk} -i\leftrightarrow j=p_\perp^2\bar{F}_{ij}$
and therefore 
\begin{equation}
\hspace{0mm} D_i A^{(1)}_j-i\leftrightarrow j~=~
g\bar{F}_{ij}
\end{equation}
so the sum 
$\bar{A}_0+\bar{A}_1$ is a pure gauge (up to $[U,V]^2$ terms).

It can be demonstrated that if we use background-Feynman gauge
for calculations in further orders in $[U,V]$ parameter we 
obtain a pure gauge field $W_i=U_i+V_i+gE_i$ where 
$(i\partial_i+g[U_i+V_i,)E^i=0$. Similarly, in the left sector the pure gauge
field at $t=-\infty$ will be ${\cal W}_i={\cal U}_i+{\cal V}_i+g{\cal E}_i$ 
with ${\cal E}$ satisfying the equation 
$(i\partial_i+g[{\cal U}_i+{\cal V}_i,){\cal E}^i=0$.

In the lowest order in $[U,V]$ we get (cf. ref. \cite{prd99,mobzor})
\begin{eqnarray}
&&\hspace{-7mm}
E_i=(x_\perp|U{p^k\over p_\perp^2}U^\dagger+V{p^k\over p_\perp^2}V^\dagger
-{p^k\over p_\perp^2}|[U_i,V_k]
-i\leftrightarrow k)
\nonumber\\
&&\hspace{-7mm}
{\cal E}_i=(x_\perp|{\cal U}{p^k\over p_\perp^2}{\cal U}^\dagger+
{\cal V}{p^k\over p_\perp^2}{\cal V}^\dagger
-{p^k\over p_\perp^2}|[{\cal U}_i,{\cal V}_k]-i\leftrightarrow k)
\label{Es}
\end{eqnarray}
The explicit expression for the correction $E_i$ in the second order in 
$[U,V]$ is obtained in the Appendix \ref{sect:ei2}.

\subsection{Gluon fields in the first order}

To find the amplitude of particle production due to the scattering of the two shock
waves we should study the behavior of gluon fields at $t\rightarrow \infty$.
The general expression for the gluon fields up to the first order in $[U,V]$ is given
in Eq. (\ref{1order}) and the corresponding diagrams are shown in Fig. \ref{fig:5}. 

Let us find the gauge field $A_\bullet^{(1)}$ in the $x_\ast,x_\bullet>0$ 
sector of the space (we use the background-Feynman gauge).
This field is a sum two terms: due to $[U_i,V^i]$ and due
to $[U_i,V_k]-i\leftrightarrow k$. The  $[U_i,V^i]$ 
term comes from the diagrams shown in  Fig. \ref{fig:5}c and \ref{fig:5}d. 
Since the Green functions in the shock-wave background in the forward cone 
$x_\ast,x_\bullet,y_\ast,y_\bullet\geq 0$ are just the bare propagators 
(\ref{barepropagators}), we obtain
\begin{eqnarray}
A_\bullet(x)&=&
g(x|{(\alpha+i\epsilon)^{-1}\over p^2+i\epsilon}|0,[U_i,V^i])\nonumber\\
&+&ig(x|{\theta(-\alpha)\over \alpha}2\pi \delta(p^2)|0,[{\cal U}_i,{\cal V}^i])
\label{1kyc}
\end{eqnarray}
where the first term in the r.h.s. of this equation comes from the diagram in 
Fig. \ref{fig:5}c and the second one from Fig. \ref{fig:5}d.
The factor $1/\alpha$ in the denominators comes from the integration over the 
final point $z_\bullet$
from $0$ to $\infty$ in the expression for the $[U_i,V^i]$ part of the linear term
(\ref{Ts}).

The term $\sim[U_i,V_k]$ is obtained by integration of the Green functions (A1)-(A3) 
in the diagrams in Fig.\ref{fig:5}b and Fig.\ref{fig:5}e 
over the final points $z_\ast,z_\perp<0$:
\begin{widetext}
\begin{eqnarray}
&&\hspace{-3mm}
A_\bullet(x)~=~\label{2kyc}\\
&&\hspace{-3mm}g{s^2\over 2}\!\int\!{d\alpha d\beta d\beta'\over 8\pi^3}
{e^{-i\alpha x_\bullet -i\beta x_\ast}\over(\alpha-i\epsilon)(\beta'-i\epsilon)}
\Big\{(x_\perp|{1\over \alpha\beta s-p_\perp^2+i\epsilon}
\partial_iU{p^k\over \alpha\beta's-p_\perp^2+i\epsilon}
U^\dagger|^{ab}0,[U_i,V_k]-i\leftrightarrow k)
\nonumber\\
&&\hspace{-3mm}
+~(x_\perp|\theta(-\alpha)2\pi\delta(\alpha\beta s-p_\perp^2)
\partial_iU{p^k\over \alpha\beta's-p_\perp^2-i\epsilon}
U^\dagger|^{ab}0,[{\cal U}_i,{\cal V}_k]-i\leftrightarrow k)
\nonumber\\
&&\hspace{-3mm}
=~g(x|{(\alpha-i\epsilon)^{-1}\over p^2+i\epsilon}|0,[U_i,E^i])
+g(x|{\theta(-\alpha)\over \alpha}2\pi\delta(p^2)|0,[{\cal U}_i,{\cal E}^i])
\nonumber
\end{eqnarray}
where we have used the notations (\ref{Es}) for brevity.
The field $A_\bullet^{1)}$ is the sum of Eqs. (\ref{1kyc}) and (\ref{2kyc}).
The field $A_\ast^{1)}$ is obtained from $A_\bullet^{1)}$ by the
replacements $x_\ast\leftrightarrow x_\bullet$ and $U\leftrightarrow V$.

Finally, let us calculate the field $A_i^{(1)}$ coming from the diagrams in Figs. 
\ref{fig:5}b and \ref{fig:5}e. Using the cluster expansion 
(\ref{cluster}) for the Green functions ${1\over P^2}P_k$ 
and the set of the propagators from Appendix \ref{sect:props},
we obtain 
\begin{eqnarray}
&&\hspace{-3mm}
A_i(x)~=~\label{Ai}\\
&&\hspace{-3mm}g\!\int\!{d\alpha d\beta \over 4\pi^2}
e^{-i\alpha x_\bullet -i\beta x_\ast}\Bigg\{
(x_\perp|{1\over \alpha\beta s-p_\perp^2+i\epsilon}
\Bigg[is\!\int\!{d\beta'\over 2\pi(\beta'-i\epsilon)}
U{p^k\over\alpha\beta' s-p_\perp^2+i\epsilon}U^\dagger
\nonumber\\
&&\hspace{-3mm}+is\!\int\!{d\alpha'\over 2\pi(\alpha'-i\epsilon)}
V{p^k\over\alpha'\beta s-p_\perp^2+i\epsilon}V^\dagger
-{1\over(\alpha-i\epsilon)(\beta-i\epsilon)}
{p^k\over\alpha\beta s-p_\perp^2+i\epsilon}\Bigg]^{ab}
|0,[U_i,V_k]-i\leftrightarrow k)
\nonumber\\
&&\hspace{-3mm}
+i(x_\perp|\theta(-\alpha)2\pi\delta(\alpha\beta s-p_\perp^2)
\Bigg[is\!\int\!{d\beta'\over 2\pi(\beta'-i\epsilon)}
{\cal U}{p^k\over\alpha\beta' s-p_\perp^2+i\epsilon}{\cal U}^\dagger
\nonumber\\
&&\hspace{-3mm}+is\!\int\!{d\alpha'\over 2\pi(\alpha'-i\epsilon)}
{\cal V}{p^k\over\alpha'\beta s-p_\perp^2+i\epsilon}{\cal V}^\dagger
-{1\over(\alpha-i\epsilon)(\beta-i\epsilon)}
{p^k\over\alpha\beta s-p_\perp^2+i\epsilon}\Bigg]^{ab}
|0,[{\cal U}_i,{\cal V}_k]-i\leftrightarrow k)\Bigg\}
\nonumber\\
&&\hspace{-3mm}=~2g(x|{1\over p^2+i\epsilon}|0,E_i)+
2ig(x|\theta(-p_0)\delta(p^2)|0,{\cal E}_i)
\nonumber
\end{eqnarray}
The left-sector fileds ${\cal A}_\mu$ are obtained by trivial replacements.

Let us present the the final set of gauge fields 
(at $x_\ast,x_\bullet>0$)
\begin{eqnarray}
&&\hspace{-0mm}
A_\bullet^{(1)}~=~
g(x|{(\alpha+i\epsilon)^{-1}\over p^2+i\epsilon}|0,[U_i,V^i])
+2g(x|{(\alpha-i\epsilon)^{-1}\over p^2+i\epsilon}|0,[U_i,E^i])
+ig(x|{\theta(-\alpha)\over \alpha}2\pi\delta(p^2)
|0,[{\cal U}_i,{\cal V}_i+2{\cal E}^i])
\nonumber\\
&&A_\ast^{(1)}~=~
-g(x|{(\beta+i\epsilon)^{-1}\over p^2+i\epsilon}|0,[U_i,V^i])
+2g(x|{(\beta-i\epsilon)^{-1}\over p^2+i\epsilon}|0,[V_i,E^i])
+ig(x|{\theta(-\beta)\over \beta}2\pi\delta(p^2)
|0,[{\cal V}_i,{\cal U}_i+2{\cal E}^i])
\nonumber\\
&&\hspace{-0mm}
A_i^{(1)}~=~2g(x|{1\over p^2+i\epsilon}|0,E_i)+
2ig(x|\theta(-p_0)2\pi\delta(p^2)|0,{\cal E}_i)
\nonumber\\
&&\hspace{-0mm}
{\cal A}_\bullet^{(1)}~=~
g(x|{(\alpha+i\epsilon)^{-1}\over p^2-i\epsilon}|0,[{\cal U}_i,{\cal V}^i])
+2g(x|{(\alpha-i\epsilon)^{-1}\over p^2-i\epsilon}|0,[{\cal U}_i,{\cal E}^i])
-ig(x|{\theta(\alpha)\over \alpha}2\pi\delta(p^2)
|0,[U_i,V_i+2E^i])
\nonumber\\
&&\hspace{-0mm}{\cal A}_\ast^{(1)}~=~
-g(x|{(\beta+i\epsilon)^{-1}\over p^2-i\epsilon}|0,[{\cal U}_i,{\cal V}^i])
+2g(x|{(\beta-i\epsilon)^{-1}\over p^2-i\epsilon}|0,[{\cal V}_i,{\cal E}^i])
-ig(x|{\theta(\beta)\over \beta}2\pi\delta(p^2)
|0,[V_i,U_i+2E^i])
\nonumber\\
&&\hspace{-0mm}
{\cal A}_i^{(1)}~=~2g(x|{1\over p^2-i\epsilon}|0,{\cal E}_i)-
2ig(x|\theta(p_0)2\pi\delta(p^2)|0,E_i)
\label{fiilds}
\end{eqnarray}
\end{widetext}
where $E$ and ${\cal E}$ are given by expressions ({\ref{Es}). 
It is easy to see that the fields (\ref{fiilds}) satisfy the boundary conditions
${\cal A}(x_\ast+x_\bullet\rightarrow\infty)=
A(x_\ast+x_\bullet\rightarrow\infty)$.
The fields
$A_\mu$ for the right sector only (with ${\cal U}_i={\cal V}_i=0$) coincide with
refs. \cite{prd99,mobzor}

\subsection{Particle production and Lipatov vertex}

The particle production is determined by the behavior of the fields at $t=\infty$
described by the Lipatov vertex
\begin{eqnarray}
&&\hspace{-0mm}
R_{\mu}(k_\perp,\eta)=\lim_{k^2\rightarrow 0}k^2A_\mu(k)
\label{lvertex}\\
&&\hspace{-0mm}=~
2gE_\mu+{2p_{2\mu}\over\alpha s}g[U_i,V^i+2E^i]+{2p_{1\mu}\over\beta s}g[V_i,U^i+2E^i]
\nonumber\\
&&\hspace{-0mm}
{\cal R}_{\mu}(k_\perp,\eta)=\lim_{k^2\rightarrow 0}k^2{\cal A}_\mu(k)
\nonumber\\
&&\hspace{-0mm}=~
2g{\cal E}_\mu+{2p_{2\mu}\over\alpha s}g[{\cal U}_i,{\cal V}^i+2{\cal E}^i]+
{2p_{1\mu}\over\beta s}g[{\cal V}_i,{\cal U}^i+2{\cal E}^i]
\nonumber
\end{eqnarray}
The vertex (\ref{lvertex}) is transverse: 
$k_{\mu}R^{\mu}(k)~=~k_iE^i+[U_i+V_i,E^i]=0$ due to our choice of $E_i$ such that
$(i\partial_i+[U_i+V_i,)E^i=0$. Similarly, $k^\mu{\cal R}_\mu(k)=0$

It is worth noting that in the case of two weak sources 
$U_i\sim (k-k')_i$ and $V_i\sim k'_i$
the vertex (\ref{lvertex}) reduces to
\begin{eqnarray}
&&\hspace{-4mm}
R_{\mu}(k;k')~\sim~
\nonumber\\
&&\hspace{-4mm}
{2\over s}(k',k-k')_\perp\Big({p_1\over\beta}-{p_2\over\alpha}\Big)_\mu
+2k'_{\perp\mu}-2k_{\perp\mu}{(k,k')_\perp\over k_\perp^2}
\nonumber\\
&&\hspace{-4mm}
=~(2k'-k)^\perp_\mu+\Big(\alpha-{2(k-k')_\perp^2\over\beta s}\Big)p_{1\mu}
-\Big(\beta-{2{k'}_\perp^2\over\alpha s}\Big)p_{2\mu}
\nonumber\\
&&\hspace{50mm}+~
{(k-k')_\perp^2-{k'}_\perp^2\over k_\perp^2}k_\mu 
\nonumber
\end{eqnarray}
where the first three terms form a standard Feynman-gauge Lipatov vertex and 
the last term is a longitudinal contribution which drops from the particle production 
amplitudes.

We get
\begin{eqnarray}
&&\hspace{0mm} 
R(k;U,V,{\cal U},{\cal V})
\label{numpart}\\
&&\hspace{0mm}
=~
-{\cal R}_\mu^a(k) R^{\mu a}(-k)~e^{~i\!\int\! d^2z_\perp 
(-{\cal U}^{ai}_z{\cal V}^{ai}_z+ U^{ai}_zV^{ai}_z)}
\nonumber
\end{eqnarray}
It is covenient to rewrite this product in terms of the transverse part of the 
Lipatov vertex in the axial $p_2^\mu A_\mu=0$ gauge: 
\begin{equation}
{\cal R}_\mu^{\rm ax}(k)=
\Big(\delta_\mu^\xi-{2p_\mu p_2^\xi\over\alpha s}\Big){\cal R}_\xi(k)
\equiv L_{\mu\perp}(k_\perp)-p_{2\mu}L(k_\perp)
\label{lvaxial}
\end{equation}
It is easy to see that $ {\cal R}_\mu R^\mu={\cal L}_iL^i$ so the 
Eq. (\ref{numpart}) reduces to
\begin{eqnarray}
&&\hspace{0mm} 
R(k;U,V,{\cal U},{\cal V})
\label{numpart1}\\
&&\hspace{0mm}
=~
{\cal L}_i^a(k_\perp) L_i^a(-k_\perp)~e^{~i\!\int\! d^2z_\perp 
(-{\cal U}^{ai}_z{\cal V}^{ai}_z+ U^{ai}_zV^{ai}_z)}
\nonumber
\end{eqnarray}
The explicit form 
of the axial-gauge Lipatov vertex is
\begin{eqnarray}
&&\hspace{0mm} 
L_i(U,V)~=2gE_i-2ig{\partial_i\over \partial_\perp^2}[U_j+2E_j,V^j]
\label{els}\\
&&\hspace{0mm}
{\cal L}_i({\cal U},{\cal V})~=~
2g{\cal E}_i-2ig{\partial_i\over \partial_\perp^2}
[{\cal U}_j+2{\cal E}_j,{\cal V}^j]
\nonumber
\end{eqnarray}
The apparent  asymmertry between $U$ and $V$ in the 
expression (\ref{els}) for $L_i$ does not affect the results - 
alternatively, one can use the expressions (\ref{els}) with 
$U\leftrightarrow V,{\cal U}\leftrightarrow {\cal V}$ and the product
(\ref{numpart1}) will remain the same.

The longitudial part $L$ can be easily obtained from (\ref{lvertex}) but we will not
need it.
Note that the dependense on $\eta$ is governed by the slope of Wilson lines.

\section{ $k_T$ factorization for the deep inelastic scattering from the nucleus}

In this section we will reproduce the standard $k_\perp$-factorized result 
\cite{mabraun1,kovtuchin,xapkovtuchin,saclay}
 for the
number of produced particles (\ref{dubfuna}) for the case when the $J_A$ is small
(e.g. virtual photon) and $J_B$ corresponds to nucleus.

A weak source  $J_A$ produces only one gluon and ${\cal J}_A$ absorbs this gluon so the upper
part of the diagram is attached to the lower by two-gluon exchange only. For the 
two-gluon exchange,
factorization formula (\ref{faktor}) simplifies to
\begin{eqnarray}
&&\hspace{-5mm}
\int\!D{\cal A} DA {\cal J}(p_A){\cal J}(p_B)e^{-iS({\cal A})+iS(A)}
\label{fafor}\\
&&\hspace{10mm}
\times~e^{i\!\int\! d^2z_\perp {\cal V}_i^a(z_\perp)~{\cal U}^{ai}(z_\perp)
-i\!\int\! d^2z_\perp  V_i^a(z_\perp)~U^{ai}(z_\perp)}
\nonumber\\
&&\hspace{-5mm}=~{1\over N_c^2-1}
\int\!D{\cal A} DA {\cal J}(p_A){\cal J}(p_B)~e^{-iS({\cal A})+iS(A)}
\nonumber\\
&&\hspace{5mm}
\times~{1\over g^4}\!
\int\! d^2z_\perp d^2z'_\perp ~
{\rm tr}\{(V^\dagger_z V_{z'}-1)({\cal V}^\dagger_{z'}{\cal V}_z-1)\}
\nonumber\\
&&\hspace{15mm}
\times~{\partial^2\over \partial z_\perp^2}~{\partial^2\over \partial{z'}_\perp^2}
{\rm tr}\{
 (U_zU^\dagger_{z'}-1)({\cal U}_{z'}{\cal U}^\dagger_z-1)\}
\nonumber 
\end{eqnarray}
where $V_z=[\infty n_1+z_\perp,-\infty n_1+z_\perp]$, cf. Eq. (\ref{faktor}). 
\begin{figure}
\includegraphics{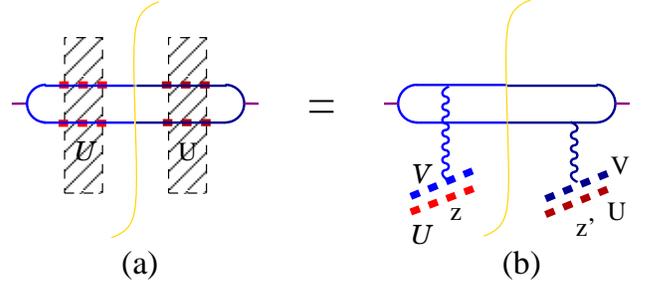}
\caption{\label{fig:3}Factorization for the two-gluon exchange.}
\end{figure}
This formula is easily seen from Fig. (\ref{fig:3}). 
Indeed, in the leading order in perturbation theory 
the l.h.s. of Eq. (\ref{fafor})
is represented by the diagram in Fig. \ref{fig:3}a 
where the quark line in the external field ${\cal U}_i,U_i$ is 
given by 
\begin{equation}
\hspace{-0mm}
\langle\chi(x)\bar{\psi}(y)\rangle~=\int\! \!dz\delta(z_\ast)
{\cal U}_x(x|{1\over \not\! p }|z)
\not\! p_2{\cal U}^\dagger_zU_z(z|{1\over \not\! p }|y)U^\dagger_y
\label{kvgreenfun}
\end{equation}
at $x_\ast,y_\ast<0$ (cf. Eq. (\ref{greenfun})).
On the other hand, the r.h.s. of Eq. (\ref{fafor}) is represented by the
diagram in Fig. \ref{fig:3}b. It is easy to see that 
the factor $\partial_\perp^2$
cancels the propagator ${1\over p^2}={1\over p_\perp^2}$ so the diagram in 
Fig. \ref{fig:3}b reduces to \ref{fig:3}a.

If we neglect the evolution, the slope of the Wilson lines $U$ can be replaced by $p_1$ 
and the slope of $V$'s by $p_2$ (see the discussion in Sec. \ref{sec:2step}).
The number of particles produced in a collision of weak and strong shock waves
(\ref{mastegral}) is given by the square of the Lipatov vertex (\ref{lvertex}).
Technically it is convenient to introduce a  source $\lambda_i(x_\perp)$ 
for the Lipatov vertex $L_i$ ( and $\sigma(x_\perp)$ for ${\cal L}$) so
\begin{eqnarray}
&&\hspace{-3mm} 
R(k;U,V,{\cal U},{\cal V})~=~\int\!d^2x_\perp d^2y_\perp~e^{i(k,x-y)_\perp}
~{\delta^2\over\delta\lambda^a_{ix}\delta\sigma^{ia}_y}
\nonumber\\
&&\hspace{-3mm}
\times~e^{i\!\int\! d^2z_\perp 
 [{\cal U}^a_i{\cal V}^{ai}+\sigma^{ia}(z_\perp){\cal L}_i^a(z_\perp)
 -U^{ai}_z V^{ai}_z-\lambda^{ia}(z_\perp)L_i^a(z_\perp)]}
\nonumber
\end{eqnarray}
For a weak source  $V_i\sim\partial_iV$ we get
$gE_i\simeq -(U(g_{ik}+{p_ip_k\over p_\perp^2})U^\dagger)^{ab}V_k^b$ so 
$L_i(x_\perp)\simeq 2(x_\perp|[{p_i\over p_\perp^2},U]p^kU^\dagger)^{ab}|V^b_k)$ 
and the above equation can be rewritten as
\begin{eqnarray}
&&\hspace{-4mm} 
R(k;U,V,{\cal U},{\cal V})~=
\int\!d^2x_\perp d^2y_\perp~e^{i(k,x-y)_\perp}
\label{rodin}\\
&&\hspace{-4mm}
\times~{\delta^2\over\delta\lambda^a_{i}(x_\perp)\delta\sigma^{ia}(y_\perp)}
~e^{i\!\int\! d^2z_\perp 
 [\tilde{\cal U}^a_i(z_\perp){\cal V}^{ai}(z_\perp) -
\tilde{U}^a_i(z_\perp) V^{ai}(z_\perp)]}
\nonumber
\end{eqnarray}
where
\begin{eqnarray}
&&\hspace{0mm}
\tilde{U}~=~e^{2{\partial_i\over \partial_\perp^2}\lambda}U
e^{-2{\partial^i\over \partial_\perp^2}(U^\dagger\lambda U)}
\nonumber\\
&&\hspace{0mm}
\tilde{\cal U}~=~e^{2{\partial_i\over \partial_\perp^2}\sigma}{\cal U}
e^{-2{\partial^i\over \partial_\perp^2}({\cal U}^\dagger\sigma {\cal U})}
\label{tildeus1}
\end{eqnarray}
(This expansion is similar to the 
functional-integral representation of the non-linear evolution as 
$e^{i\phi_1}Ue^{-i\phi_2}$ 
developed in ref. \cite{pl}).
We need here only the first two terms of the expansion in powers of 
$\lambda$ and $\sigma$.

Using the simplified factorization formula (\ref{fafor}) we get

\begin{eqnarray}
&&\hspace{-5mm} 
R(k;U,V,{\cal U},{\cal V})\nonumber\\
&&\hspace{-3mm}=~{1\over g^4}\!\int\!d^2x_\perp d^2y_\perp d^2z_\perp d^2z'_\perp~
{e^{i(k,x-y)_\perp}\over N_c^2-1}
{\delta^2\over\delta\lambda^a_{ix}\delta\sigma^{ia}_y}
\nonumber\\
&&\hspace{-5mm}\times
 ~{\rm tr}\{
\tilde{\cal U}^\dagger_z \tilde{U}_z\tilde{U}^\dagger_{z'}\tilde{\cal U}_{z'}\}
\partial_{\perp z}^2~\partial_{\perp z'}^2
{\rm tr}\{(V^\dagger_zV_{z'}-1)
({\cal V}^\dagger_{z'}{\cal V}_z-1)\}
\nonumber\\
&&\hspace{-5mm}
=~{4g^{-2}\over(N_c^2-1)}\int\! d^2x_\perp d^2y_\perp d^2z_\perp d^2z'_\perp ~
\nonumber\\
&&\hspace{-5mm}
\times~{\rm tr}\big\{\big[(y|{p_i\over p_\perp^2}|z)
({\cal U}_{z'}{\cal U}^\dagger_z t^a
-~{\cal U}_{z'}{\cal U}^\dagger_yt^a{\cal U}_y{\cal U}^\dagger_{z'})
\nonumber\\
&&\hspace{13mm}
-~(y|{p_i\over p_\perp^2}|z')(t^a{\cal U}_{z'}{\cal U}^\dagger_z 
-{\cal U}_{z'}{\cal U}^\dagger_yt^a{\cal U}_y{\cal U}^\dagger_z)\big]
\nonumber\\
&&\hspace{-5mm}
\times~\big[(x|{p^i\over p_\perp^2}|z)(t^aU_zU^\dagger_{z'}
-U_zU^\dagger_xt^aU_xU^\dagger_{z'})
\nonumber\\
&&\hspace{13mm}
-~(x|{p^i\over p_\perp^2}|z')(U_zU^\dagger_{z'}t^a
-U_zU^\dagger_xt^aU_xU^\dagger_{z'})\big]\big\}\nonumber\\
&&\hspace{-5mm}
\times
~e^{i(k,x-y)_\perp}
\partial_{\perp z}^2~\partial_{\perp z'}^2
{\rm tr}\{(V^\dagger_zV_{z'}-1)
({\cal V}^\dagger_{z'}{\cal V}_z-1)\}
\label{rdva}
\end{eqnarray}
Without evolution, ${\cal U}\equiv U$ and ${\cal V}\equiv V$
so the square of the Lipatov vertex (\ref{rdva}) reduces to

\begin{eqnarray}
&&\hspace{-3mm}
R(k_\perp;U,V)~\label{erkauv}\\
&&\hspace{-3mm}=~{2g^{-2}\over N_c^2-1}\!\int\!d^2z_\perp d^2z'_\perp~
{\rm tr}\{\partial_\perp^2V_z~\partial_\perp^2 V^\dagger_{z'}+z\leftrightarrow z'\}
\nonumber\\
&&\hspace{-3mm}\times
 ~{\rm Tr}\big\{\big((k_\perp|[{p_i\over p_\perp^2},U]|z_\perp)-
\big((k_\perp|[{p_i\over p_\perp^2},U]|z'_\perp)\big)
\nonumber\\
&&\hspace{-3mm}\times~
\big((z_\perp|[{p^i\over p_\perp^2},U^\dagger]|k_\perp)
-(z'_\perp|[{p^i\over p_\perp^2},U^\dagger]|k_\perp)\big)\big\}
\nonumber
\end{eqnarray}
It is easy to see that in the momentum representation we get 
\begin{eqnarray}
&&\hspace{-5mm}
R(k_\perp;U,V)~=~{2g^{-2}\over N_c^2-1}\!\int\!d^2k^\perp_1 d^2k^\perp_2~
K^{\rm emi}_{\rm BFKL}(k_1,k_2;k)
\nonumber\\
&&\hspace{-5mm}
\times~{\rm tr}\{V(k^\perp_1) V^\dagger(-k^\perp_2)\}
{\rm tr}\{U(k_\perp-k^\perp_1 )U^\dagger(k^\perp_2-k^\perp)\}
\label{rklass}
\end{eqnarray}
where 
\begin{eqnarray}
&&\hspace{-3mm}
K^{\rm emi}_{\rm BFKL}(k_1,k_2;k)=-2\Big({k^i\over k_\perp^2}k_{1\perp}^2-k_1^i\Big)
\Big({k_i\over k_\perp^2}k_{2\perp}^2-k_{2i}\Big)
\nonumber\\
&&\hspace{-3mm}
=~{k_{1\perp}^2(k-k_2)_\perp^2\over k_\perp^2}
+{k_{2\perp}^2(k-k_1)_\perp^2\over k_\perp^2}-(k_1-k_2)_\perp^2
\end{eqnarray}
is the gluon-emission part of the BFKL kernel \cite{bfkl}.

Substituting the result of the integration over central-rapidity gluons (\ref{erkauv})
into the factorization formula (\ref{enge}) we obtain the standard $k_T$-factorization 
formula
\begin{eqnarray}
n_g(k_\perp;\eta))&=&{2g^{-2}\over N_c^2-1}\!\int\!\!d^2k_1 d^2k_2
K^{\rm emi}_{\rm BFKL}(k_1,k_2;k_\perp)
\nonumber\\
&\times&
\langle A|{\rm Tr}\{V(k_1)V^\dagger(-k_2)|A \rangle
\nonumber\\
&\times&
\langle B|{\rm Tr}\{U(k_\perp-k_1)U^\dagger(k_\perp-k_2)|B 
 \rangle
 \label{katifak}
\end{eqnarray}
where the rapidity dependence comes from the slope of the Wilson lines in the r.h.s.
This formula was obtained in the approximation when
we neglect the evolution of the shock waves, but it can be proved without this
assumption \cite{mabraun1,kovtuchin,xapkovtuchin,saclay}.
It should be emphasized that Eq. (\ref{katifak}) is valid only in the first order in 
the $[U,V]$ expansion (that is, for the $pA$ scattering). As we shall see below, it is not valid
beyond the first order.
 
\section{\label{secteffect}Effective action}

In this section we get the first-order effective action for the
double functional integral ({\ref{doubfun}) and check that the
Lipatov vertex (\ref{els}) serves as a ``splitting function''
for the non-linear evolution equation \cite{npb96,yura}
\begin{eqnarray}
&&{\partial\over\partial\eta}{\rm Tr}\{U_xU^\dagger_y\}~=~-
{\alpha_s\over 4\pi}\!\int\! d^2z_\perp{(x-y)_\perp^2\over(x-z)_\perp^2(y-z)_\perp^2}
\nonumber\\
&&\times
({\rm Tr}\{U_xU^\dagger_z\}{\rm Tr}\{U_zU^\dagger_y\}-N_c{\rm Tr}\{U_xU^\dagger_y\})
\label{bk}
\end{eqnarray}
The effective action is defined by the functional integral (\ref{mastegral})
without $A_\mu(p)A^\mu(-p)$ insertion:
\begin{eqnarray}
&&\hspace{-4mm} 
e^{iS_{\rm eff}(U,V,{\cal U},{\cal V})}
~=~
\int\!\!D{\cal A} DA~\exp\Big\{-iS({\cal A})
\label{effactdef}\\
&&\hspace{-4mm}
 +~i\!\int\! d^2z \Big({\cal V}^{ai}_z
 [\infty,{\cal F}_{\ast i},\infty]^a_z
+~(\infty,{\cal F}_{\ast i},\infty)^a_z
 {\cal U}^{ai}_z\Big)
\nonumber\\
&&\hspace{-4mm}
+~iS(A)-i\!\!\int\! \!d^2z \Big(
 V^{ai}_z 
 [\infty,F_{\ast i},\infty]^a_z
  + 
 (\infty,F_{\ast i},\infty)^a_z
 U^{ai}_z\Big)\Big\}
 \nonumber
\end{eqnarray}
Performing the shift  $A\rightarrow A+\bar{A}_i^{(0)}$, 
${\cal A}\rightarrow {\cal A}+\bar{\cal A}_i^{(0)}$ we get
(cf. Eq. (\ref{mastegral1})
\begin{eqnarray}
&&\hspace{-3mm} 
e^{iS_{\rm eff}(U,V,{\cal U},{\cal V})}~=~
\int\!D{\cal A} DA
 \exp\Big\{i\!\int\! d^4z\Big(
({1\over 2}{\cal A}^\mu\bar{\cal D}_{\mu\nu}{\cal A}^\nu 
  \nonumber\\
&&\hspace{-3mm}-~
 {1\over 2} A^\mu \bar{D}_{\mu\nu}A^\nu
 -{\cal T}^\mu{\cal A}_\mu+T^\mu A_\mu\Big)\Big\}
 \label{mastegrale1}
\end{eqnarray}
The effective action in the first nontrivial order is 
given by the integration of linear terms with the appropriate Green functions
\begin{eqnarray}
&&\hspace{0mm} 
iS_{\rm eff}(U,V,{\cal U},{\cal V})
~=~
\nonumber\\
&&\hspace{0mm}
{1\over 2}\!\int\! d^4zd^4z'
\Big\{-T^a_\mu(z)\langle A^{\mu a}(z) A^{\nu b}(z')\rangle T^b_\nu(z')
 \nonumber\\
&&\hspace{20mm}+~2T^a_\mu(z)
\langle A^{\mu a}(z) {\cal A}^{\nu b}(z')\rangle {\cal T}^b_\nu(z')
  \nonumber\\
&&\hspace{20mm}-~{\cal T}^a_\mu(z)\langle {\cal A}^{\mu a}(z) 
{\cal A}^{\nu b}(z')\rangle {\cal T}^b_\nu(z')\Big\}
  \nonumber
 \label{seffgen}
\end{eqnarray}
The first term here can be taken from \cite{prd99,mobzor}:
\begin{eqnarray}
&&\hspace{-5mm} 
iS^{(1)}_{\rm eff}(U,V)~=
\label{effgen1}\\
&&\hspace{-5mm}
\alpha_s\Delta\eta\int\! d^2z\Big(E^a_iE^{ai}+
[U_i,V^i+2E^i]{1\over \partial^2_\perp}[V_i,U^i+2E^i]\Big) 
\nonumber
\end{eqnarray}
where $\Delta\eta=\eta_1-\eta_2$ is the difference of rapisdities of the 
slopes of $U$ and $V$ Wilson lines.
The third term $S^{(1)}_{\rm eff}({\cal U},{\cal V})$ in Eq. (\ref{seffgen}) 
is obtained from Eq. (\ref{effgen1}) by
the usual replacements $U\leftrightarrow \cal{U}$, $V\leftrightarrow \cal{V}$

Let us calculate the second term beginning with the contribution 
$\sim [U_i,V^i]...[{\cal U}^i,{\cal V}^i]$.
 The Green function in the $x_\ast,x_\bullet>0$ sector is simply
 the perturbative propagator (\ref{barepropagators}) so one obtains
\begin{eqnarray}
&&\hspace{-3mm} 
-g^2\!\!\int\!{d\alpha d\beta\over 2\pi^2\alpha\beta}
(0,[{\cal U}^i,{\cal V}_i]^a|2\pi\delta(\alpha\beta s-p^2_\perp)
\theta(\alpha)|0,[U^i,V_i]^a)
\nonumber\\
&&\hspace{-3mm}=~-g^2
([U^i,V_i]^a|{1\over p_\perp^2}[U^i,V_i]^a)\int_0^\infty\!{d\alpha \over \pi\alpha}
\label{effgen1a}
\end{eqnarray}
The integral (\ref{effgen1a}) is formally divegrent at both small and large
$\alpha$. This divergence occurs because we've put the slopes of Wilson lines 
$e_1$ and $e_2$ (see Eq. (\ref{eqn20}))
to be $p_1$ and $p_2$. If we keep the slopes off the light cone, we get 
(see ref. \cite{mobzor}):
\begin{eqnarray}
&&\hspace{0mm} 
-g^2\!\int\!{d\alpha d\beta\over 2\pi^2}
{1\over(\alpha+e^{-\eta_1}\beta)(\beta+e^{\eta_2}\alpha)}
\nonumber\\
&&\hspace{0mm}\times~
(0,[{\cal U}^i,{\cal V}_i]^a|2\pi\delta(\alpha\beta s
-p^2_\perp)\theta(\alpha)|0,[U^i,V_i]^a)~\nonumber\\
&&\hspace{0mm}
=~-2\alpha_s\Delta\eta([{\cal U}^i,{\cal V}_i]^a|{1\over p_\perp^2}[U^i,V_i]^a)
\label{withslope}
\end{eqnarray}
The contributions of $[U_i,V_k]-i\leftrightarrow k$ terms can be calculated in
a similar manner by integration of these terms with appropriate Green functions
in Eq. (\ref{effgen1}). After some algebra, one obtains
\begin{eqnarray}
&&\hspace{-5mm} 
-2\alpha_s\Delta\eta\Big\{\!\int\! d^2z_\perp {\cal E}^iE_i(z_\perp)
+({\cal U}^i-{\cal V}_i,{\cal E}^i]^a|{1\over p_\perp^2}|[U^i,V_i]^a)
\nonumber\\
&&\hspace{-5mm}
+~([{\cal U}^i,{\cal V}_i]^a|{1\over p_\perp^2}|[U^i-V_i,E^i]^a)\Big\}
\label{effgen1b}
\end{eqnarray}
The sum of
Eqs. (\ref{withslope}) and (\ref{effgen1b}) give
the second term in the effective action
\begin{eqnarray}
&&\hspace{-3mm} 
iS^{(2)}_{\rm eff}(U,V,{\cal U},{\cal V})~=
\nonumber\\
&&\hspace{-3mm}
-\alpha_s\Delta\eta\int\! d^2z\Big(2{\cal E}^{ai}E^a_i+
[{\cal U}_i,{\cal V}^i+2{\cal E}^i]{1\over \partial^2_\perp}[V_i,U^i+2E^i]
\nonumber\\
&&\hspace{-3mm}+~ [{\cal V}_i,{\cal U}^i
+{\cal E}^i]{1\over \partial^2_\perp}[U_i,V^i+E^i]
\Big) 
\label{effgen2}
\end{eqnarray}
It is easy to see that the total effective action 
$S_{\rm eff}(U,V,{\cal U},{\cal V})=S^{(1)}_{\rm eff}(U,V)+
S^{(2)}_{\rm eff}(U,V,{\cal U},{\cal V})+
S^{(3)}_{\rm eff}({\cal U},{\cal V})$ can  be represented as
\begin{eqnarray}
&&\hspace{0mm} 
iS_{\rm eff}(U,V,{\cal U},{\cal V})~=~-i{\cal U}_i{\cal V}^i
+iU_iV^i
\label{effact}\\
&&\hspace{0mm}
+{\alpha_s\over 4}\Delta\eta\int\! d^2z\Big(L^a_iL^{ai}-2{\cal L}^{ai}L^a_i
+{\cal L}^{ai}{\cal L}^a_i\Big)
\nonumber
\end{eqnarray}
where $L_i$ and ${\cal L}_i$ are given by Eq. (\ref{els}).
Note that the effective action is a square of Lipatov vertex (\ref{lvertex}):
$R_\mu R^\mu=L_iL^i$, $ {\cal R}_\mu R^\mu={\cal L}_iL^i$, 
${\cal R}_\mu {\cal R}^\mu={\cal L}_i{\cal L}^i$.

For the future applications we will rewrite the effective action
(\ref{effact}) as a Gaussian integration over the auxiliary field 
$\lambda$ coupled to Lipatov vertex (\ref{els}):
\begin{eqnarray}
&&\hspace{0mm} 
e^{iS_{\rm eff}(U,V,{\cal U},{\cal V})}
~=~e^{i\int d^2z_\perp(-{\cal U}_i{\cal V}^i
+U_iV^i)}
\label{effacta}\\
&&\hspace{0mm}\times~\int D\lambda 
\exp\Big\{\int\! d^2z_\perp
\Big(-{\lambda^a_i\lambda^{ai}\over \alpha_s\Delta\eta}
+(L^a_i-{\cal L}^{ai})\lambda^a_i\Big)\Big\}
\nonumber
\end{eqnarray}
A single
field $\lambda_i$ for both $L_i$ and ${\cal L}_i$ reflects 
the fact that gauge fields
$A_\mu$ and ${\cal A}_\mu$ coincide at $t=\infty$.

Let us prove now that the effective action (\ref{effacta}) agrees with the 
non-linear evolution equation. 
 To find the evolution of the dipole $U_xU^\dagger_y$, we need to consider
the effective action for the weak source $V$. As we mentioned above, at small 
$V_i\sim\partial_iV$ one has
$gE_i\simeq -(U(g_{ik}+{p_ip_k\over p_\perp^2})U^\dagger)^{ab}V_k^b$ so 
$L_i(x_\perp)\simeq 2(x_\perp|[{p_i\over p_\perp^2},U]p^kU^\dagger)^{ab}|V^b_k)$ 
and Eq. (\ref{effacta}) can be rewritten as
\begin{eqnarray}
&&\hspace{-5mm} 
\int\!D{\cal A} DA 
 \exp\Big\{-iS({\cal A})+iS(A)
\label{effactu}\\
&&\hspace{-5mm}-~i\!\int\! d^2z_\perp 
 {\cal V}^{ai}_z
 [\infty,{\cal F}_{\ast i},\infty]^a_z
-i\!\int\! d^2z_\perp 
 (\infty,{\cal F}_{\ast i},\infty)^a_z
 {\cal U}^{ai}_z
\nonumber\\
&&\hspace{-5mm} 
+~i\!\int\! d^2z_\perp 
 V^{ai}_z
 [\infty,F_{\ast i},\infty]^a_z
  +i\!\int\! d^2z_\perp 
 (\infty,F_{\ast i},\infty)^a_z
 U^{ai}_z\Big\}\nonumber\\
&&\hspace{-5mm} 
=\int D\lambda 
\exp\Big\{\int\! d^2z_\perp\Big(-{\lambda^a_i\lambda^{ai}\over 
\alpha_s\Delta\eta}
-i{\cal V}_i\tilde{\cal U}^i+iV_i\tilde{U}^i\Big)\Big\}
  \nonumber
\end{eqnarray}
where
\begin{eqnarray}
&&\hspace{0mm}
\tilde{U}~=~e^{2{\partial_i\over \partial_\perp^2}\lambda}U
e^{-2{\partial^i\over \partial_\perp^2}(U^\dagger\lambda U)}
\nonumber\\
&&\hspace{0mm}
\tilde{\cal U}~=~e^{2{\partial_i\over \partial_\perp^2}\lambda}{\cal U}
e^{-2{\partial^i\over \partial_\perp^2}({\cal U}^\dagger\lambda {\cal U})}
\label{tildeus}
\end{eqnarray}
Again,we need here only the terms up to the second order in $\lambda$. 
\footnote{Strictly speaking,
to get the effective action (\ref{effacta}) we need only the first term of the 
expansion of the exponents in Eq. (\ref{tildeus}) in powers of $\lambda$. However, in
order to reproduce the full non-linear equation (\ref{bk}) we need the 
gluon-reggeization terms coming
from the second order in expansion in $\lambda$. Formally the gluon reggeization 
exceeds the accuracy of the semiclassical calculation
of the effective action; however, when $V_i$ is not large the gluon reggeization 
is of the same order as (\ref{effacta}).}

 To find the
evolution of the dipole $U_xU^\dagger_y$ we should expand Eq. 
(\ref{effactu}) in powers of $V_i({\cal V}_i)$ and use the 
formula 
$(U_xU^\dagger_y)^{\eta_2}=Pe^{ig\!\int_x^y\!dz_i[\infty,F_{\ast i},\infty]_z}$.
We get
\begin{eqnarray}
&&\hspace{0mm}
(U_xU^\dagger_y{\cal U}_y{\cal U}^\dagger_x)^{\eta_1}
\nonumber\\
&&\hspace{0mm}~=\int D\lambda ~
e^{-{1\over \alpha_s\Delta\eta}\int\! d^2z_\perp\lambda^a_i\lambda^{ai}}~
\tilde{U}_x\tilde{U}^\dagger_y\tilde{\cal U}_y\tilde{\cal U}^\dagger_y
\end{eqnarray}
Performing the Gaussian integration over $\lambda$ one obtains after some algebra
\begin{eqnarray}
&&\hspace{-4mm}
{\rm tr}(U_xU^\dagger_y{\cal U}_y{\cal U}^\dagger_x)^{\eta_1}
\label{difbk}\\
&&\hspace{-4mm}=~
{\rm tr}\{U_xU^\dagger_y{\cal U}_y{\cal U}^\dagger_x\}^{\eta_2}
+{\alpha_s\Delta\eta\over 4\pi^2}\!\int\! d^2z_\perp 
{(x-y)_\perp^2\over(x-z)_\perp^2(z-y)_\perp^2}
\nonumber\\
&&\hspace{-4mm}
\times~({\rm tr}\{{\cal U}^\dagger_xU_xU^\dagger_z{\cal U}_z\}
{\rm tr}\{{\cal U}^\dagger_zU_zU^\dagger_y{\cal U}_y\}
-N_c{\rm tr}\{{\cal U}^\dagger_xU_xU^\dagger_y{\cal U}_y\})^{\eta_2}
\nonumber
\end{eqnarray}
which is the non-linear evolution equation for the double 
functional integral for the cross section \cite{difope,mobzor}.

\section{\label{sect7}Classical fields and Lipatov vertex in the second order}

If we neglect the evolution, the classical  sources $U_i$ and ${\cal U}_i$
 coincide (and similarly $V_i={\cal V}_i$) so the corresponding fields 
in the right and left sectors coincide and 
are determined by the integrals of the retarded Green functions with 
the ${\cal T}=T$ sources.  

\subsection{First-order gluon field in the $x_\parallel^\mu A_\mu=0$ gauge}

\begin{figure}
\includegraphics{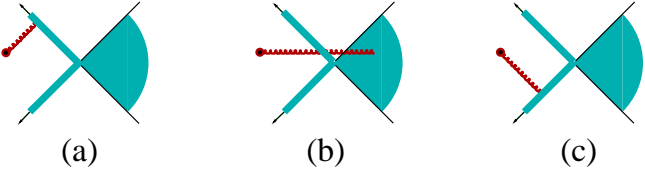}
\caption{\label{fig:1or}Retarded classical field in the first order}
\end{figure}
%
 In the first order in $[U,V]$, the classical
fields (\ref{fiilds}) (in the background gauge) reduce to
\begin{eqnarray}
\hspace{-3mm}
{\bar A}_i^{(1)}&=&W_i(x_\perp)\theta(-x_\ast)\theta(-x_\bullet)
+~ U_i(x_\perp)\theta(-x_\ast)\theta(x_\bullet)
\nonumber\\
&+&V_i(x_\perp)\theta(x_\ast)\theta(-x_\bullet)
+2g(x|{1\over p^2+i\epsilon p_0}|0,E_i)
\nonumber\\
\bar{A}_\bullet^{(1)}&=&
g(x|{(\alpha+i\epsilon)^{-1}\over p^2+i\epsilon p_0}|0,[U_i,V^i])
\nonumber\\
&~&\hspace{25mm}+~g(x|{(\alpha-i\epsilon)^{-1}\over p^2+i\epsilon p_0}|0,[U_i,E^i])
\nonumber\\
\bar{A}_\ast^{(1)}&=&
-g(x|{(\beta+i\epsilon)^{-1}\over p^2+i\epsilon p_0}|0,[U_i,V^i])
\nonumber\\
&~&\hspace{25mm}+~
g(x|{(\beta-i\epsilon)^{-1}\over p^2+i\epsilon p_0}|0,[V_i,E^i])
\nonumber\\
\bar{A}_i^{(1)}&=&2g(x|{1\over p^2+i\epsilon p_0}|0,E_i)
\label{klfiilds}
\end{eqnarray}
We see that the fields outside the forward cone 
are piece-wise pure gauge:
\begin{eqnarray}
&&\hspace{-0mm}
{\bar A}_i^{(1)}~=~W_i(x_\perp)\theta(-x_\ast)\theta(-x_\bullet)
+~ U_i(x_\perp)\theta(-x_\ast)\theta(x_\bullet)
\nonumber\\
&&\hspace{20mm}
+~V_i(x_\perp)\theta(x_\ast)\theta(-x_\bullet)
\nonumber\\
&&\hspace{-0mm}
\bar{A}_\bullet^{(1)}~=~-ig\delta(x_\ast)\theta(-x_\bullet)
(x_\perp|{1\over p_\perp^2}|[U_i,E^i]),
\nonumber\\
&&\hspace{-0mm}
\bar{A}_\ast^{(1)}~=~-ig\delta(x_\bullet)\theta(-x_\ast)
(x_\perp|{1\over p_\perp^2}|[V_i,E^i])
\label{outside}
\end{eqnarray}
while the field in the  forward sector 
$x_\ast,x_\bullet>0$ is determined by the Lipatov vertex (\ref{lvertex}).
\begin{eqnarray}
&&\hspace{-0mm}
\bar{A}_\mu^{(1)}~=~(x|{1\over p^2+i\epsilon p_0}|R_\mu)
\label{inside}
\end{eqnarray}
%
\begin{figure*}
\includegraphics{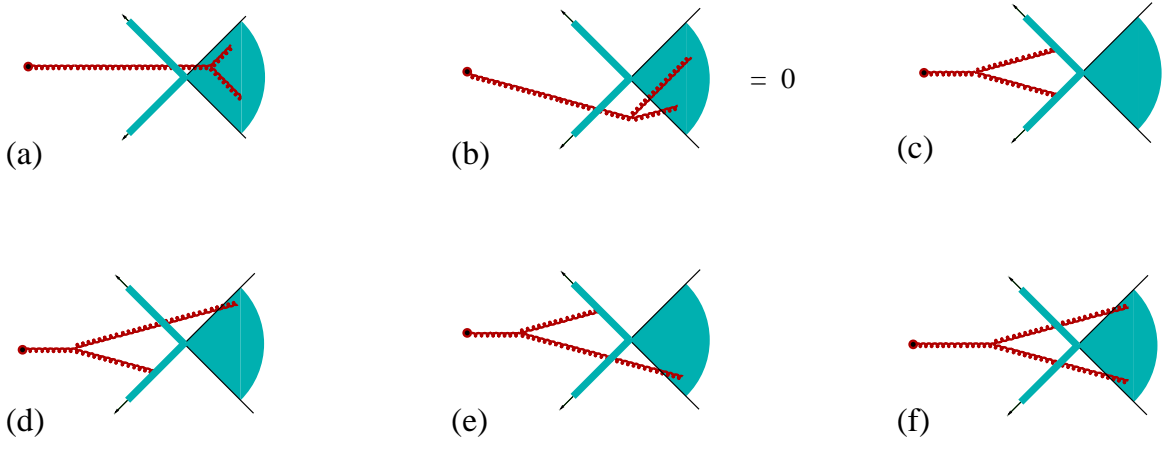}
\caption{\label{fig:6}Classical field in the second order in $[U,V]$}
\end{figure*}
%
Summarising, the classical field (\ref{klfiilds})  can be represented as
\begin{eqnarray}
{\bar A}_\mu^{(1)}&=&
\tilde{A}_{\perp\mu}
+(x|{1\over p^2+i\epsilon p_0}|R^{(1)}_\mu)
\label{vezde1}
\end{eqnarray}
where
\begin{eqnarray}
\tilde{A}_i&=& \theta(-x_\ast)\theta(x_\bullet)U_i(x_\perp)\label{tildaa}\\
&+&\theta(x_\ast)\theta(-x_\bullet)V_i(x_\perp)+
\theta(-x_\ast)\theta(-x_\bullet)W_i(x_\perp)
\nonumber
\end{eqnarray}
is a trivial part corresponding to a piece-wise pure-gauge field and
\begin{eqnarray}
&&\hspace{-3mm}
R^{(1)}_\mu=2gE^\perp_\mu
+g{2p_{2\mu}\over s}\Big(
{1\over\alpha +i\epsilon}[U_i,V^i]+{2\over\alpha -i\epsilon}[U_i,E^i]\Big)
\nonumber\\
&&\hspace{-3mm}+~g
{2p_{1\mu}\over s}\Big(
{1\over\beta +i\epsilon}[V_i,U^i]+{2\over\beta +i\epsilon}[V_i,E^i]\Big)
\label{rmu}
\end{eqnarray}
describes the non-trivial part related to the gluon emission.

The axial-gauge Lipatov vertex is given by first line in Eq. (\ref{els}):
\begin{eqnarray}
&&\hspace{0mm} 
L_i(U,V)~=~2gE_i-2ig{\partial_i\over \partial_\perp^2}[U_j+2E_j,V^j]
\label{el}
\end{eqnarray}
Following ref. \cite{nncoll}, it is instructive 
to represent the fields (\ref{klfiilds}) in the 
gauge $x^\parallel_\mu A^\mu=x_\ast A_\bullet+x_\bullet A_\ast=0$. 
In this gauge one obtains at $x_\ast, x_\bullet>0$
\begin{eqnarray}
&&\hspace{-0mm}
A^\parallel_\mu(x_\parallel,x_\perp)~=~-\int_1^\infty\! udu ~x_\parallel^\rho
F_{\rho\mu}(ux_\parallel,x_\perp),\nonumber\\
&&\hspace{-0mm}
A^\perp_\mu(x_\parallel,x_\perp)~=~-\int_1^\infty\! du ~x_\parallel^\rho 
F_{\rho\mu}(ux_\parallel,x_\perp)
\label{mvgage}
\end{eqnarray}
where $x_\parallel\equiv {2\over s}x_\ast p_1+{2\over s}x_\bullet p_2$
(so that $x=x_\parallel +x_\perp$).
The limit of integration $\infty$ in the above expressions 
was chosen to satisfy our boundary condition (no pure gauge fields
at $t=\infty$ 
\footnote{The requirement of absence of pure gauge fileds at $t=\infty$ differs 
from the condition 
$A(0,x_\perp)=0$ adopted in the papers \cite{nncoll} and 
therefore the fields (\ref{mvfiilds}) differ from ref. \cite{nncoll}.
For the same reason, the boundary condition $\alpha_{1,2}=\alpha_\perp$ 
from ref. \cite{nncoll} is not 
satisfied by our fields (\ref{mvfiilds})}.

The fields ouside the forward cone (\ref{outside})
 trivially satisfy the $x^\parallel_\mu A^\mu=0$ gauge condition. 
 The fields in the forward cone 
$x_\ast,x_\bullet>0$ are obtained
by integrating $F_{\mu\nu}$'s corresponding to the fileds in the bF gauge 
(\ref{klfiilds}).
From the Eq. (\ref{mvgage}) we get:
\begin{eqnarray}
&&\hspace{-3mm}
\bar{A}_\ast(x)~=~i{gs\over 8\pi x_\bullet}
\!\int\!d^2z_\perp
\theta(-x_\parallel^2+(x-z)_\perp^2)
\nonumber\\ 
&&\hspace{20mm}
\times~(2[U_i,V^i]+
[U_i-V_i,E^i])
\label{mvfiilds}\\ 
&&\hspace{-3mm}
\bar{A}_\bullet^{\rm MV}(x)~=~-i{gs\over 8\pi x_\ast}
\!\int\!d^2z_\perp
\theta(-x_\parallel^2+(x-z)_\perp^2)
\nonumber\\ 
&&\hspace{20mm}
\times~(2[U_i,V^i]+
[U_i-V_i,E^i]) 
\nonumber\\ 
&&\hspace{-3mm}
\bar{A}_i(x)~=~-{g\over\pi}\!\int\!d^2z_\perp
\delta(x_\parallel^2-(x-z)_\perp^2)E_i
\nonumber\\ 
&&\hspace{-3mm}+
~g\!\int\!d^2z_\perp
\theta(-x_\parallel^2+(x-z)_\perp^2)
(x_\perp|{p_i\over p_\perp^2}|z_\perp)[U_j+V_j,E^j]
\nonumber
\end{eqnarray}
Similarly to ref. \cite{nncoll}, the fields  
$x_\ast A^{(1)}_\bullet$, $x_\bullet A^{(1)}_\ast$ and 
$A^{(1)}_i$ are boost invariant.
However, as we mentioned in the footnote, the fields (\ref{mvfiilds}) differ from those
in ref. \cite{nncoll} due to a different boundary condition.

\subsection{Gluon field and Lipatov vertex in the second order in $[U,V]$}

In the next order the classical field $\bar{A}^{(2)}$ is given by 
diagrams in Fig. \ref{fig:6} calculated in the Appendix C. The result of
the calculation is given by the sum of the piece-wise pure gauge field 
and the field of the gluon emission described by the second-order 
Lipatov vertex represented by two terms coming from the diagrams in 
Fig. \ref{fig:1or} and Fig. \ref{fig:6}
\begin{eqnarray}
\hspace{-7mm}{\bar A}_i&=&\theta(-x_\ast)\theta(-x_\bullet)W_i(x_\perp)
\label{vezde}\\
\hspace{-7mm}
&+& \theta(-x_\ast)\theta(x_\bullet)U_i(x_\perp)
+\theta(x_\ast)\theta(-x_\bullet)V_i(x_\perp)
\nonumber\\
\hspace{-7mm}
&+&(x|{1\over p^2+i\epsilon p_0}|R^{(1)}_\mu+R^{(2)}_\mu)~
+~O([U,V]^3)
\nonumber
\end{eqnarray}
The first part of the Lipatov vertex coming from the diagrams in Fig. \ref{fig:1or}
and Fig. \ref{fig:6}a
has the form
\begin{eqnarray}
&&\hspace{-0mm}
R^{\rm (1)}_\mu(k)~=~R^{\rm (1)}_{\perp\mu}(k_\perp)+
{2p_{1\mu}\over s}\Big(
{R_{1+}^{\rm (1)}(k_\perp)\over \beta+i\epsilon}
+{R_{1-}^{\rm (1)}(k_\perp)\over \beta-i\epsilon}\Big)
\nonumber\\
&&\hspace{-0mm}+~
{2p_{2\mu}\over s}\Big({R_{2+}^{\rm (1)}(k_\perp)\over \alpha+i\epsilon}
+{R_{2-}^{\rm (1)}(k_\perp)\over \alpha-i\epsilon}\Big)
\label{R(1)}
\end{eqnarray}
where the notations are
\begin{eqnarray}
&&\hspace{-5mm}
(R^{\rm (1)}_{\perp\mu})^a~=~2gE_{\perp\mu}+
4ig((\partial^\perp_\mu U){1\over p_\perp^2}U^\dagger)^{ab}[V_i,E^i]^b
\nonumber\\
&&\hspace{25mm}+~
4ig((\partial^\perp_\mu V){1\over p_\perp^2}V^\dagger)^{ab}[U_i,E^i]^b
\nonumber\\
&&\hspace{-5mm}
(R_{1+}^{\rm (1)})^a~=~-g[U_i,V^i]^a-
2g((\partial_\perp^2V){1\over p_\perp^2}V^\dagger)^{ab}[U_i,E^i]^b
\nonumber\\
&&\hspace{-5mm}
(R_{1-}^{\rm (1)})^a~=~2gp_\perp^2(U{1\over p_\perp^2}U^\dagger)^{ab}
[V_i,E^i]^b
\nonumber\\
&&\hspace{-5mm}
(R_{2+}^{\rm (1)})^a~=~g[U_i,V^i]^a-
2g((\partial_\perp^2U){1\over p_\perp^2}U^\dagger)^{ab}[V_i,E^i]^b
\nonumber\\
&&\hspace{-5mm}
(R_{2-}^{\rm (1)})^a~=~
2gp_\perp^2(V{1\over p_\perp^2}V^\dagger)^{ab}[U_i,E^i]^b
\label{R(1)s}
\end{eqnarray}
The second-order term coming from the diagrams in Fig. \ref{fig:6}c-f is given by
\begin{eqnarray}
&&\hspace{-3mm}
R^{\rm (2)~a}_{\mu}(k)
~=
~f^{abc}\!\int\! {d^2k'_\perp\over 32\pi^2}\Bigg\{-{1\over \sqrt{\cal G}}[K_\mu g_{\mu \eta}
-~2(k_\xi g_{\mu \eta}\nonumber\\
&&\hspace{-3mm}
-~(k_\eta g_{\mu \xi})] 
\tilde{R}^{b\xi}(k_\perp)\tilde{R}^c_{\xi}(k_\perp-k'_\perp)
+8i\Big({p_2\over \alpha s}-{p_1\over \beta s}\Big)_{\mu}
\nonumber\\
&&\hspace{-3mm}
\times~{(k',k-k')_\perp+{k^2\over 2}-i\sqrt{\cal G}\over {k'}_\perp^2(k-k')_\perp^2}
R_1^b(k'_\perp) R_2^c(k_\perp-k'_\perp) 
 \nonumber\\
&&\hspace{-3mm}
+~32i{k^\xi\delta_\mu^\eta-k^\eta\delta_\mu^\xi\over {k'}_\perp^2}
\Big({p_{1\xi}\over\beta s}R_{1-}^b(k'_\perp)
+{p_{2\xi}\over\alpha s}R_{2-}^b(k'_\perp)\Big) 
\nonumber\\
&&\hspace{-3mm}
\times~gE_\eta^{\perp c}(k_\perp-k'_\perp)\Bigg\}
\label{R(2)}\\
&&\hspace{-3mm}
+~g^2\Big({4p_1\over \beta s}-
~{4p_2\over \alpha s}\Big)_{\mu}
(k_\perp|[U_i,V^i]^{ab}{1\over p_\perp^2}|[U_j+V_j,E^j]^b)
\nonumber
\end{eqnarray}
where we use the notations 
\begin{equation}
K\equiv (k-2k')_\perp+{k^\parallel\over \alpha\beta s}(k,k-2k')_\perp
\label{K}
\end{equation}
\begin{equation}
{\cal G}~=~{k'}_\perp^2(k-k')_\perp^2-\Big({k^2+i\epsilon k_0\over 2}+
(k',k-k')_\perp)\Big)^2
\label{calje}
\end{equation}
%
\begin{eqnarray}
&&\hspace{-1mm}
\tilde{R}_\xi(k'_\perp)=2E_{\perp\xi}(k'_\perp)
\label{Rs}\\
&&\hspace{-1mm} +~{(k,k')_\perp+{k^2\over 2}-i\sqrt{\cal G}\over {k'}_\perp^2}
\Big({2p_1\over \beta s}R_1(k'_\perp)+{2p_2\over \alpha s}R_2(k'_\perp)\Big)_\xi
\nonumber
\end{eqnarray}
and $R_1^{\rm (1)}\equiv R_{1+}^{\rm (1)}+R_{1-}^{\rm (1)}$,
$R_2^{\rm (1)}\equiv R_{2+}^{\rm (1)}+R_{2-}^{\rm (1)}$.\\ 
With $[U,V]^2$ accuracy
$R_1^{\rm (1)}$ and $R_2^{\rm (1)}$ in Eq. (\ref{R(2)}) can be simplified to
\begin{eqnarray}
&&\hspace{-3mm}
(R_1^{\rm (1)})^a~=~-g[U_i,V^i]^a+2gp_\perp^2(U{1\over p_\perp^2}U^\dagger)^{ab}
[V_i,E^i]^b
\nonumber\\
&&\hspace{-3mm}
(R_2^{\rm (1)})^a~=~g[U_i,V^i]^a
+2gp_\perp^2(V{1\over p_\perp^2}V^\dagger)^{ab}[U_i,E^i]^b
\label{Rypr}
\end{eqnarray}
The second-order Lipatov vertex is the the sum of Eqs. (\ref{R(1)}) and
(\ref{R(2)}) at the mass shell $k^2=0$. At the first sight, it looks like
 the expression (\ref{R(2)}) is divergent
at $k'_\perp\parallel k_\perp$ since ${\cal G}=k_\perp^2{k'}^2(1-\cos^2\theta)$.
This collinear divergence is however purely longitudinal and therefore
can be eliminated by proper gauge transformation. To see that, let us write 
the Lipatov vertex in the
axial light-like gauge $p_2^\mu A_\mu=0$ (\ref{lvaxial}). As we mentioned above,
 only the first transverse term in r.h.s. of Eq. (\ref{lvaxial}) 
is essential since $p_2$ term does not contribute to the square of the Lipatov
vertex. For this transverse part we obtain
\begin{eqnarray}
&&\hspace{-3mm}
L_i(k_\perp)~=~L^{\rm (1)}_i(k_\perp)+L^{\rm (2)}_i(k_\perp)~
\label{Lvertex}\\
&&\hspace{-3mm}
L^{\rm (1)}_i(k_\perp)~=~R^{\rm (1)}_i(k_\perp)+
{2k_i\over k_\perp^2}
(R_{1+}^{\rm (1)}(k_\perp)+R_{1-}^{\rm (1)}(k_\perp))
\nonumber\\
&&\hspace{-3mm}
L^{\rm (2)~a}_i(k_\perp)
~=~f^{abc}\!\int\! {d^2k'_\perp\over 4\pi^2}
{1\over \sqrt{k_\perp^2{k'}_\perp^2-(k,k')_\perp^2}}
\label{lvertex2}\\
&&\hspace{-3mm}
\times~
\Bigg\{{1\over 8}\Big(g_{ij}+{k_ik_j\over k_\perp^2}\Big)(2k'-k)^j
L^{(1)b}_l(k'_\perp)L^{(1)cl}(k_\perp-k'_\perp)
\nonumber\\
&&\hspace{-3mm}
+~{1\over 2}(k_j -{(k,k')_\perp\over {k'}_\perp^2
}k'_j)L^{(1)bj}(k'_\perp)L^{(1)c}_i(k_\perp-k'_\perp)\Bigg\}
\nonumber\\
&&\hspace{-3mm}+~if^{abc}\!\int\! {d^2k'_\perp\over 4\pi^2{k'}_\perp^2}
\Bigg[2[U_j+V_j,E^j]^b(k'_\perp)E_i^c(k_\perp-k'_\perp)
\nonumber\\
&&\hspace{-3mm}+~2{k_ik^j\over k_\perp^2}[U_l-2E_l,V^l]^b(k'_\perp)E_j^c(k_\perp-k'_\perp)
\nonumber\\
&&\hspace{-3mm}-~
\Big({k'_i\over (k-k')_\perp^2}
+{k_i\over k_\perp^2}\Big)(R_1^{\rm (1)})^b(k'_\perp)
(R_2^{\rm (1)})^c(k_\perp-k'_\perp)\Bigg]
\nonumber\\
&&\hspace{-3mm}
-~4{k_i\over k_\perp^2}(k_\perp|[U_i,V^i]^{ab}{1\over p_\perp^2}|[U_j+V_j,E^j]^b)
\nonumber
\end{eqnarray}
Note that since
the square bracket in the r.h.s. of the above equation vanishes at
$k'_\perp\parallel k_\perp$, the collinear divergrnce is absent.
The second-order Lipatov vertex (\ref{Lvertex}) is the main technical result
of this paper.

\section{Conclusions and outlook}

Let us summarize the progress towards the solution of the main problem  - 
the particles/fields produced in the collision of two shock waves. 
The Yang-Mills equations 
with sources $U$ and $V$ describe the two shock waves corresponding to the colliding
hadrons. The expansion of the classical fields in commutators $[U,V]$ has the advantage
of being ``symmetric'' in contrast to the  usual expansion in powers of the strength of one 
of the sources.  We have calculated the second nontrivial term of the expansion. 
This term  is relevant for the description of 
$dA$ scattering, similar to the first term $\sim [U,V]$ describing the $pA$ collisions.
\begin{figure}
\includegraphics{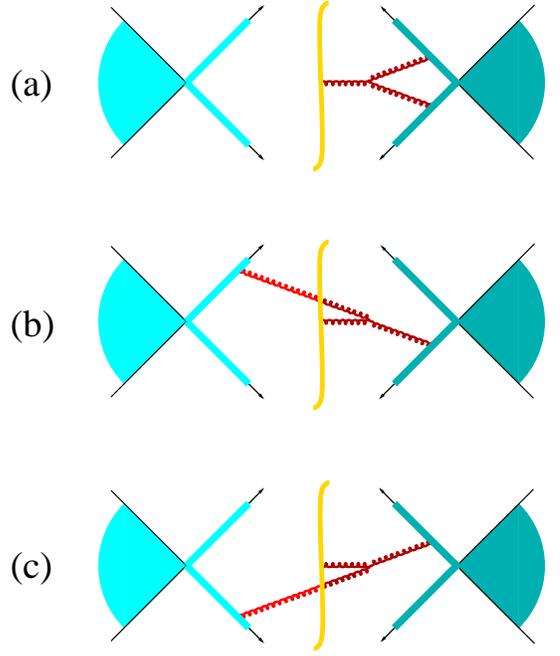}
\caption{\label{fig:7}Classical field for ${\cal U}\neq U$, ${\cal V}\neq V$, }
\end{figure}
\newpage

Note that while the first-order field given by Eq. (\ref{vezde1}) (or Eq. (\ref{mvfiilds}))
is real, the second-order field has an imaginary part given by the second term in braces in Eq. 
(\ref{R(2)}). The real part of the second-order term is given by Eq. (\ref{R(1)}) plus 
the first expression in braces in Eq. (\ref{R(2)}) represented by the product of first-order Lipatov 
vertices. I think that this univeral structure will survive to the higher orders of the commutator
expansion. Unfortunately, the explicit form of the imaginary part of the field 
(second term in the r.h.s of the Eq. (\ref{lvertex2})) does not suggest any idea 
how this expression may look in higher orders in $[U,V]$ expansion. 
Technically, the relative simplicity of the real part is a consequence of its relation 
to the leading log approximation (LLA).
If we consider the general case ${\cal U}\neq U$, ${\cal V}\neq V$, the second-order 
classical field would contain $\Delta\eta$ just like the effective action (\ref{effact}).
Indeed,  if we calculate the field $A_\mu$ in the right sector, the typical 
expression  
$$
\int\! {du \over(u+i\epsilon)Z(u)}[U,V](k')[U,V](k-k')
$$
(see Eq. (\ref{2kycok})) would be replaced by
\begin{eqnarray}
&&\hspace{-2mm}
\int_0^1\! {du \over(u+i\epsilon)Z(u)}[U,V](k')[U,V](k-k')
\nonumber\\
&&\hspace{-2mm}+~
\int_{-\infty}^0\! {du \over(u+i\epsilon)Z(u)}[{\cal U},{\cal V}](k')[U,V](k-k')+
\nonumber\\
&&\hspace{-2mm}+~\int_1^\infty\! {du \over(u+i\epsilon)Z(u)}[U,V](k')[{\cal U},{\cal V}](k-k')
\end{eqnarray}
where the three terms correspond to the contributions of diagram in Fig. \ref{fig:7}a,b,
and c, respectively.
 The integration over $u$ is regularized by the width of the shock wave 
 (cf.{\cite{mobzor,balbel}))
and only the real part is survives - the imaginary part exceeds the accuracy of the LLA.

If we consider the amplitude rather than the cross section, we  
take only one set of fields (to the right of the cut) and impose the usual Feynman boundary 
conditions. In this case the classical field $A_\mu$ 
is the sum of logarithms of the type $(\alpha_s\ln s)^n[U,V]^n$ (cf. \cite{prd99}).
The imaginary part calculated above 
may be related to an old idea due to Lipatov that one can unitarize the BFKL pomeron 
if one finds the proper $i\pi$ or $-i\pi$ to each $\ln s$ in the LLA approximation. Indeed, 
 both these imaginary parts come from one source - causality:
the $i\pi$'s in the amplitude come from the dispersion relations based on causality, while 
$i\pi$'s in the classical field (\ref{R(2)}) come from retarded propagators.

\begin{acknowledgments}
The author thanks F. Gelis, E. Iancu, Yu. Kovchegov and 
R. Venugopalan for valuable discussions. The author is grateful to 
theory groups at CEA Saclay and  LPTHE Jussieu for kind hospitality. 
This work was supported by contract
DE-AC05-84ER40150 under which the Southeastern Universities Research
Association (SURA) operates the Thomas Jefferson National Accelerator
Facility.
\end{acknowledgments}

\appendix

\section{\label{sect:a1}Green functions in a shock-wave background}

\subsection{\label{sect:props}Feynman rules for cross sections in 
a shock-wave background}

Let us present the set of the bF-gauge propagators in 
the background of a shock-wave 
field ${\cal U}_i\theta(-x_\ast),U_i\theta(-x_\ast)$ (\cite{difope}).

At $x_\ast,y_\ast>0$ all propagators are bare, see egn. (\ref{barepropagators}).

At $x_\ast>0,y_\ast<0$ we get 
\begin{eqnarray}
&&\hspace{-2mm}
\langle A_\mu(x)A_\nu(y)\rangle
~=~(x|{1\over p^2+i\epsilon }
O_{\mu\nu}(U){1\over p^2+i\epsilon }|y)U^\dagger_y
\label{greenfun1}\\
&&\hspace{-2mm}
\langle{\cal A}_\mu(x){\cal A}_\nu(y)\rangle
~=~(x|{1\over p^2-i\epsilon }
O_{\mu\nu}({\cal U}){1\over p^2-i\epsilon }|y){\cal U}^\dagger_y
\nonumber\\
&&\hspace{-2mm}
\langle{\cal A}_\mu(x)A_\nu(y)\rangle
\nonumber\\
&&\hspace{12mm}=~-i(x|2\pi\delta(p^2)\theta(p_0)
O_{\mu\nu}(U){1\over p^2+i\epsilon }|y)U^\dagger_y
\nonumber\\
&&\hspace{-2mm}
\langle A_\mu(x){\cal A}_\nu(y)\rangle
\nonumber\\
&&\hspace{12mm}=~i(x|2\pi\delta(p^2)\theta(-p_0)
O_{\mu\nu}({\cal U}){1\over p^2-i\epsilon }|y){\cal U}^\dagger_y
\nonumber
\end{eqnarray}
while at $x_\ast,y_\ast<0$ the propagators are
\begin{eqnarray}
&&\hspace{-5mm}
\langle A_\mu(x)A_\nu(y)\rangle
~=~U_x(x|{-i\over p^2+i\epsilon }|y)U^\dagger_y
\label{greenfun2}\\
&&\hspace{-5mm}
\langle{\cal A}_\mu(x){\cal A}_\nu(y)\rangle
~=~{\cal U}_x(x|{i\over p^2-i\epsilon }|y){\cal U}^\dagger_y
\nonumber\\
&&\hspace{-5mm}
\langle{\cal A}_\mu(x)A_\nu(y)\rangle
~=~-{\cal U}_x(x|{1\over p^2-i\epsilon }
O_{\mu\nu}({\cal U}^\dagger U){1\over p^2+i\epsilon }|y)U^\dagger_y
\nonumber
\end{eqnarray}

where 
\begin{eqnarray}
&&\hspace{-2mm}O_{\mu\nu}(U)~=
~\!\int\! dz \delta({2\over s}z_\ast)~|z)\Big\{2\alpha
g_{\mu\nu}U
\label{calo}\\
&&\hspace{-2mm}+~{4i\over s}(p_{2\nu}\partial_\mu U+\mu\leftrightarrow\nu)
-{4p_{2\mu}p_{2\nu}\over \alpha s^2}\partial_\perp^2U\Big\}(z|
\nonumber
\end{eqnarray}

\subsection{Retarded propagators}

First, let us present the retarded propagator in the background of the shock wave 
$\theta(x_\ast)\Omega_i(x_\perp)+\theta(-x_\ast)\Lambda_i(x_\perp)$ where 
$\Omega_i=\Omega i\partial_i\Omega^\dagger$ and 
$\Lambda_i=\Lambda i\partial_i\Lambda^\dagger$ are the pure gauge fields. This 
propagator can be obtained from  Eqs. (\ref{greenfun1}), (\ref{greenfun2}) 
by setting $U={\cal U}=\Omega^\dagger\Lambda$, taking appropriate combinations, 
and rotating by the matrix $\Omega$:
\begin{eqnarray}
&&\hspace{-2mm}
\langle A(x)A(y)\rangle_{\rm ret}=\theta(x_\ast)\theta(y_\ast)
\Omega_x(x|{1\over p^2+i\epsilon p_0}|y)\Omega^\dagger_y
\nonumber\\
&&\hspace{-2mm}
+~\theta(-x_\ast)\theta(-y_\ast)
~\Lambda_x(x|{1\over p^2+i\epsilon p_0}|y)\Lambda^\dagger_y
\label{retprop1}\\
&&\hspace{-2mm}
+~\theta(x_\ast)\theta(-y_\ast)\Omega_x(x|{1\over p^2+i\epsilon p_0}
O_{\mu\nu}(\Omega^\dagger\Lambda){1\over p^2+i\epsilon p_0}|y)\Lambda^\dagger_y
\nonumber
\end{eqnarray}

\subsubsection{Cluster expansion}

The background field in our calculations is the trial configuration
$\bar{A}_i=\theta(-x_\ast)U_i+\theta(-x_\bullet)V_i$. 
Since we are expanding in powers of commutators $[U,V]$, the adequate procedure 
for the propagator in the $\bar{A}_i$ background is the cluster expansion
(\ref{cluster}):
\begin{eqnarray}
&&\Big({1\over P^2+i\epsilon p_0}\Big)_{U+V}~
\label{kluster}\\
&&=~
\Big({1\over P^2+i\epsilon p_0}\Big)_{U}
+\Big({1\over P^2+i\epsilon p_0}\Big)_{V}
-{1\over p^2+i\epsilon p_0}+...
\nonumber
\end{eqnarray}
where dots stand for the second and higher terms of cluster expansion.
Most often, the first term (\ref{kluster}) is sufficient. In several cases when 
we need the second term,
the fillowing trick helps.

Let us add and subtract $E_i$ to our trial configuration so it takes the form
$\bar{A}_i=\tilde {A}_i-E_i\theta(-x_\ast)\theta(-x_\bullet)$ where $\tilde {A}_i$ is 
the piece-wise pure-gauge
field given by Eq. (\ref{tildaa}), see Fig. \ref{fig:xz3}. 
With $[U,V]^2$ accuracy, the propagator in the $\bar{A}_i$ background takes the form:
\begin{equation}
{1\over P^2+i\epsilon p_0}={1\over\tilde{P}^2+i\epsilon p_0}+
{1\over P^2+i\epsilon p_0}\{p^i,E_i\}{1\over P^2+i\epsilon p_0}
\label{expande}
\end{equation}
where we can replace $(P^2+i\epsilon p_0)^{-1}$ in the second term in r.h.s. by
the first term in cluster expansion (\ref{kluster}). The remaining first term
in the r.h.s. of eq. (\ref{expande}) is calculated below.

\subsubsection{Propagator in the piece-wise pure-gauge field.}

The retarded propagator 
$\langle A_\mu(x)A_\nu(y)\rangle_{\rm ret}$ in the background of the 
piece-wise pure gauge configuration shown in Fig. \ref{fig:xz3}
\begin{figure}
\includegraphics{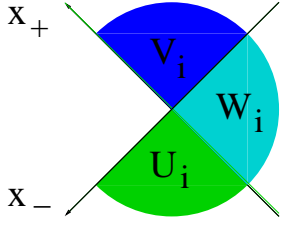}
\caption{\label{fig:xz3}Piece-wise pure-gauge field $\tilde{A}_i$.}
\end{figure}
can be obtained by ``squaring'' of the propagator in the background of one 
shock wave (\ref{retprop1}).
In the region $x_\ast>0,x_\bullet<0$ and  $y_\ast,y_\bullet<0$ it
is given by Eq, (\ref{retprop1}) with appropriate substitutions:
\begin{eqnarray}
&&\hspace{-2mm}
\langle A_\mu(x)A_\nu(y)\rangle_{\rm ret}
\label{retpropa1}\\
&&\hspace{-2mm}
=~V_x(x|{1\over p^2+i\epsilon p_0}
O_{\mu\nu}(V^\dagger W){1\over p^2+i\epsilon p_0}|y)W^\dagger_y
\nonumber
\end{eqnarray}
where
\begin{eqnarray}
&&\hspace{-2mm}O_{\mu\nu}(U)\label{bigo}\\
&&\hspace{-2mm}=~\!\int\! dz \delta({2\over s}z_\ast)~|z)
\Big\{2\alpha
g_{\mu\nu}V^\dagger W+{4i\over s}
(p_{2\nu}\partial_\mu(V^\dagger W)
\nonumber\\
&&\hspace{-2mm}+\mu\leftrightarrow\nu)
-{4p_{2\mu}p_{2\nu}\over \alpha s^2}(\partial_\perp^2(V^\dagger W)
+i[U_i,V^i]W)
\Big\}(z|
\nonumber
\end{eqnarray}
Here the last term $\sim [U_i,V^i]$ (additional in comparison to 
Eq. (\ref{calo})) is due to the source contribution to the second
variational derivative of the action
\begin{eqnarray}
&&\hspace{-2mm}
{\delta^2 S\over\delta A^a_{\mu}(x)\delta A^b_{\mu}(y)}
\label{seconder}\\
&&\hspace{-2mm}=~\Big[D^2 g_{\mu\nu}+2iF_{\mu\nu}-{4p_{2\mu}p_{2\nu}\over s^2}
\Big\{{1\over\alpha }\delta({2\over s}x_\ast)(\partial_i
\nonumber\\
&&\hspace{-2mm}
+~i[V_i)U^i
+{1\over\beta }
\delta({2\over s}x_\bullet)(\partial_i+i[U_i)V^i\Big\}\Big]^{ab}\delta(x-y)
\nonumber
\end{eqnarray}

The propagator in the region $x_\ast<0,x_\bullet>0$ and  $y_\ast,y_\bullet<0$ is
similar to Eq. (\ref{retpropa1})
\begin{eqnarray}
&&\hspace{-2mm}
\langle A_\mu(x)A_\nu(y)\rangle_{\rm ret}
\label{retpropa2}\\
&&\hspace{-2mm}
=~U_x(x|{1\over p^2+i\epsilon p_0}
\tilde{O}_{\mu\nu}(U^\dagger W){1\over p^2+i\epsilon p_0}|y)W^\dagger_y
\nonumber
\end{eqnarray}
where
\begin{eqnarray}
&&\hspace{-2mm}\tilde{O}_{\mu\nu}(U^\dagger W)\label{bigkalo}\\
&&\hspace{-2mm}=~\!\int\! dz \delta({2\over s}z_\bullet)~|z)
\Big\{2\beta
g_{\mu\nu}U^\dagger W+{4i\over s}
(p_{2\nu}\partial_\mu(U^\dagger W)
\nonumber\\
&&\hspace{-2mm}+\mu\leftrightarrow\nu)
-{4p_{2\mu}p_{2\nu}\over \beta s^2}(\partial_\perp^2(U^\dagger W)
+i[V_i,U^i]W)
\Big\}(z|
\nonumber
\end{eqnarray}

Finally, the propagator at $x_\ast,x_\bullet>0$ and  $y_\ast,y_\bullet<0$ 
has the form:
\begin{eqnarray}
&&\hspace{-2mm}
\langle A_\mu(x)A_\nu(y)\rangle_{\rm ret}\label{retpropa3}\\
&&\hspace{-2mm}=~i
(x|{1\over p^2+i\epsilon p_0}\Big\{O_{\mu\xi}(U){1\over p^2+i\epsilon p_0}
\tilde{O}^\xi_{~\nu}(U^\dagger W)
\nonumber\\
&&\hspace{-2mm}
+~i
\tilde{O}_{\mu\xi}(V)
{1\over p^2+i\epsilon p_0}
O^\xi_{~\nu}(V^\dagger W)\Big\}{1\over p^2+i\epsilon p_0}|y)W^\dagger_y
\nonumber
\end{eqnarray}

\section{\label{sect:ei2}Pure gauge field $E$ in the second order}

From $F_{\mu\nu}(U_i+V_i+E_i)=0$ we get 
\begin{equation}
(\partial_i-i[W_i,)E_j-i\leftrightarrow j~=~
i([U_i,V_j]-i\leftrightarrow j-[E_i,E_j])
\end{equation}
If we choose the $(\partial_i-i[W_i,)E^i=0$ condition the above equation
reduces to the recursion formula
\begin{equation}
E^a_i~=~-g(x_\perp|W{p^k\over p_\perp^2}W^\dagger|^{ab}
[U_i,V_k]^b-i\leftrightarrow k-g[E_i,E_k]^b)
\label{recurformula}
\end{equation}
It is convenient to introduce complex coordinates in the 2-dimensional 
plane: $z=z_1+iz_2,\bar{z}=z_1-iz_2$
and $Q=Q_1+iQ_2,\bar{Q}=Q_1-iQ_2$ for
arbitrary vector $Q$. In these
notations the recursion formula (\ref{recurformula}) simplifies to
\begin{equation}
E^a~=~-g(x_\perp|W{i\over \bar{p}}W^\dagger|^{ab}
K_{12}-E_{12})
\label{recurformula1}
\end{equation}
where $K_{12}\equiv [U_1,V_2]-[U_2,V_1]$ and $E_{12}\equiv [E_1,E_2]$.
In the leading order $W{1\over \bar{p}}W^\dagger$ can be approximated by cluster
expansion: $W{1\over \bar{p}}W^\dagger\simeq\Big({1\over \bar{P}}\Big)_{(1)}$ 
where $\Big({1\over \bar{P}_{(1)}}\Big)\equiv U{1\over \bar{p}}U^\dagger
+V{1\over \bar{p}}V^\dagger-{1\over \bar{p}}$ and therefore we get Eq. (\ref{Es}). 
In the second order we need one more term of the cluster expansion:
\begin{eqnarray}
&&W{1\over \bar{p}}W^\dagger~\simeq~\Big({1\over \bar{P}_{(1)}}\Big)
+(U{1\over \bar{p}}U^\dagger 
-{1\over \bar{p}})\bar{p}(V{1\over \bar{p}}V^\dagger -{1\over \bar{p}})
\nonumber\\
&&+~(U{1\over \bar{p}}U^\dagger 
-{1\over \bar{p}})\bar{p}(V{1\over \bar{p}}V^\dagger -{1\over \bar{p}})
-\Big({1\over \bar{P}_{(1)}}\Big)\bar{E}^{(1)}\Big({1\over \bar{P}_{(1)}}\Big)
\nonumber
\end{eqnarray}
so the second-order expression for $E$ is
\begin{eqnarray}
&&\hspace{-4mm}E^{a(2)}~=~
-ig(x_\perp|\Big({1\over P}\Big)_{(1)}|^{ab}[E^{1}_1,E^{1}_2]^b)
\label{E2}\\
&&\hspace{-4mm}-~
ig(x_\perp|(U{1\over p}U^\dagger 
-{1\over p})p(V{1\over p}V^\dagger -{1\over p})+U\leftrightarrow V
|^{ab}K^b_{12})
\nonumber
\end{eqnarray}
Similarly, for the $\bar{E}$ component we get
\begin{eqnarray}
&&\hspace{-4mm}\bar{E}^{a(2)}~=~
ig(x_\perp|\Big({1\over {\bar P}_{(1)}}\Big)|^{ab}[E^{1}_1,E^{1}_2]^b)
\label{E2c}\\
&&\hspace{-4mm}+~
ig(x_\perp|(U{1\over {\bar p}}U^\dagger 
-{1\over \bar{p}})\bar{p}(V{1\over \bar{p}}V^\dagger -{1\over \bar{p}})
+U\leftrightarrow V
|^{ab}K^b_{12})
\nonumber
\end{eqnarray}

The corresponding formula for the pure gauge
field ${\cal E}$ in the left sector is obtained by the trivial replacements
$U\rightarrow{\cal U}$ and $V\rightarrow{\cal V}$.

\section{Classical fields in the second order}
\subsection{The fields at $x_\ast,x_\bullet<0$}
\begin{figure}
\includegraphics{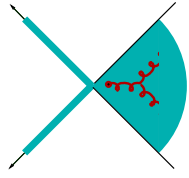}
\caption{\label{fig:xz1}Classical field in the backwards cone}
\end{figure}

Since all the Green functions in our expansion are retarded, the only 
second-order contribution order the classical field $\bar{A}^{(2)}$ is 
comes from
the $D^kF_{ik}$ part of the linear term shown in  Fig. \ref{fig:xz1}.
At $x_\ast, x_\bullet<0$ the gluons in 
Fig. \ref{fig:xz1} propagate in the external field 
$U_i+V_i$. It is convenient to add 
(and subtract later) the external field $E_i$. 
The contribution of the
diagram in Fig. \ref{fig:xz1} gives then
\begin{eqnarray}
&&\hspace{-4mm}
i\int\! d^4z (x|{\theta(-z_\ast)\theta(-z_\bullet)
\over (P-E)^2g_{ik}+2i\bar{G}_{ik}}|z)^{ab}D^jL^b_{kj}(z)
\nonumber\\
&&\hspace{-4mm}
=~-(x_\perp|W{p^k\over p_\perp^2}W^\dagger|^{ab}L_{ik}^b)
+2(x_\perp|W{p^k\over p_\perp^2}W^\dagger|^{ab}[E_i,E_k]^b)
\nonumber\\
&&\hspace{12mm}+~
(x_\perp|W{1\over p_\perp^2}W^\dagger|^{ab}[L_{ik},E^k]^b)
\nonumber
\end{eqnarray}
Since each of the two legs in the diagram in  Fig. \ref{fig:6}a 
represents $E_i$ with our accuracy, the Fig. \ref{fig:6}a contribution 
can be reduced to 
\begin{eqnarray}
&&\hspace{-2mm}
(x_\perp|W{1\over p_\perp^2}W^\dagger|^{ab}
i[D_iE_k-i\leftrightarrow k,E^k]^b)
\nonumber\\
&&\hspace{-2mm}
-~(x_\perp|W{p^k\over p_\perp^2}W^\dagger|^{ab}[E_i,E_k]^b)
\nonumber
\end{eqnarray}
Combining these two terms we get (at $x_\ast, x_\bullet,0$)
\begin{eqnarray}
&&\hspace{-2mm}
A^a_i(x)~
=~-(x_\perp|W{p^k\over p_\perp^2}W^\dagger|^{ab}L_{ik}^b)
\nonumber\\
&&\hspace{-2mm}+~(x_\perp|W{p^k\over p_\perp^2}W^\dagger|^{ab}[E_i,E_k]^b)
~=~E^a_i(x_\perp)
\end{eqnarray}
up to the terms $\sim[U,V]^3$, see Eq. (\ref{E2}). Also, it is easy to see 
that the
longitudinal components $A_\bullet$ and $A_\ast$ vanish at  $x_\ast, x_\bullet<0$.

\subsection{The fields at $x_\ast>0, x_\bullet<0$}

\begin{figure}
\includegraphics{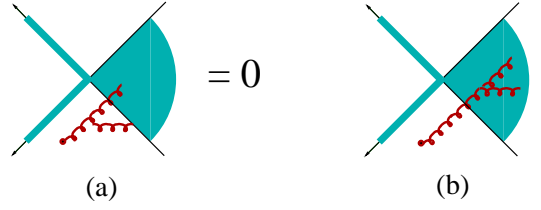}
\caption{\label{fig:xz2}Classical field iat $x_\ast>0,x_\bullet<0$}
\end{figure}

First, we  note that there are two types of diagrams shown in 
Fig. \ref{fig:xz2}: with the three-gluon vertex in the  
$z_\ast>0, z_\bullet<0$ quadrant and in the $z_\ast, z_\bullet<0$ quadrant.  
 The contribution of the first type (see  Fig. \ref{fig:xz2}a) 
 vanishes  because the only
non-zero component of the first-order field $\bar{A}^{(1)}_\mu$ 
in this case is 
$\bar{A}^{(1)}_\bullet$ such that $D_\ast\bar{A}^{(1)}_\bullet=0$, see the Eq.
(\ref{outside}).

Next we calculate the diagram in Fig. \ref{fig:xz2}b. As in the previous case,
the gluon legs are attached only to the $D^kF_{ik}$ part of the linear term 
(\ref{Ts}). 
The gluons in Fig. \ref{fig:xz2}b propagate in the external field 
$U_i\theta(-x_\ast)+V_i\theta(-x_\bullet)$. The propagator 
in this background is given by the cluster expansion (\ref{kluster}) 
or, if one needs the $[U,V]^2$ accuracy, by Eq. (\ref{expande})
and formulas(\ref{retpropa1})-(\ref{retpropa3}).

Let us start with the transverse component of the field $A_i$. 
If the three-gluon 
vertex is integrated over only  the $z_\ast,z_\bullet<0$ quadrant one can 
demonstrate that similarly to 
the $x_\ast, x_\bullet<0$ case, the contribution of the diagram in Fig. \ref{fig:xz2}b  
reduces to 
\begin{eqnarray}
&&\hspace{-2mm}
A^a_i(x)~
=~\int\! d^4z \theta(-z_\ast)\theta(-z_\bullet)
\label{aitiy}\\
&&\hspace{-0mm}
(x|{1\over \tilde {P}^2}\tilde{P}^k|z)^{ab}
\Big(L_{ik}^b(z_\perp)-[E_i,E_k]^b(z_\perp)\Big)
\nonumber
\end{eqnarray}
where $P_i=i\partial_i+g\tilde{A}_i$, 
$\tilde{A}_i=U_i\theta(-x_\ast)+V_i\theta(-x_\bullet)+
E_i\theta(-x_\ast)\theta(-x_\bullet)$. Using the Green function
in the  the two-shock-wave background (\ref{retpropa2}), we see that
the r.h.s. of Eq. (\ref{aitiy}) vanish so $A_i^{(1)}=A_i^{(2)}=0$. 

It is easy to see that $A_\ast=0$ at $x_\ast>0,x_\bullet<0$ so we are left with
$A_\bullet$ only. Again, since the only contribution from the three-gluon vertex
comes from the  $z_\ast,z_\bullet<0$ cone, it can be demonstrated that 
\begin{eqnarray}
&&\hspace{-2mm}
A^a_\bullet(x)~
=~-2i\!\int\! d^4z \theta(-z_\ast)\theta(-z_\bullet)
\label{abullet}\\
&&\hspace{-0mm}
\times~(x|{1\over \tilde {P}^2}\bar{F}_{\bullet i}
{1\over \tilde {P}^2}\tilde{P}^k|z)^{ab}
\Big(L_{ik}(z_\perp)-[E_i,E_k](z_\perp)\Big)^b
\nonumber
\end{eqnarray}
Substituting the explicit form of the Green function 
(\ref{greenfun1},\ref{greenfun2}) one obtains
\begin{eqnarray}
&&\hspace{-0mm}
-2i\!\int\! d^4z \theta(-z_\ast)\theta(-z_\bullet)
(x|V{1\over p^2}V^\dagger\bar{F}_{\bullet i}W{p^k\over p^2}W^\dagger|z)^{ab}
\nonumber\\
&&\hspace{-0mm}
\times~\Big(L_{ik}(z_\perp)-[E_i,E_k](z_\perp)\Big)^b
\nonumber\\
&&\hspace{-0mm}
=~2(x|V{1\over p^2}V^\dagger{1\over\alpha-i\epsilon}|^{ab}0,[U_i,E^i]^b)
\delta({2\over s}x_\ast)
\nonumber
\end{eqnarray}
Thus, the only non-vanishing component of the classical filed in the 
$x_\ast\geq 0,x_\bullet<0$ region is
\begin{eqnarray}
&&\hspace{-10mm}
A^a_\bullet(x)
~=~-2i\delta({2\over s}x_\ast)(x_\perp|V{1\over p_\perp^2}V^\dagger|^{ab}
[U_i,E^i]^b)
\end{eqnarray}
Similarly, at $x_\ast<0,x_\bullet\geq 0$ 
\begin{eqnarray}
&&\hspace{-10mm}
A^a_\bullet(x)~=~-2i\delta({2\over s}x_\bullet)
(x_\perp|U{1\over p_\perp^2}U^\dagger|^{ab}[V_i,E^i]^b)
\end{eqnarray}
\subsection{The fields in the forward cone $x_\ast,x_\bullet>0$}

\subsubsection{The three-gluon vertex in the backward cone}

At first, we consider the contribution from Fig. \ref{fig:6}a 
where the three-gluon vertex integrated over  
$z_\ast, z_\bullet<0$ quadrant. This contribution is similar to the one 
we considered in the previous section so we can start with the expressions
(\ref{aitiy}) and (\ref{abullet}). Using the Green functions in the 
$\tilde{A}$ 
background given by Eq. (\ref{expande})
and formulas (\ref{retpropa1})-(\ref{retpropa3}), one obtains
\begin{eqnarray}
&&\hspace{-5mm}
\bar{A}_\mu^a(x)=\int\! d^4z (x|{1\over p^2+i\epsilon p_0}\Big\{2|0,E^a_{\perp\mu})
\\
&&\hspace{-5mm}
+~{4\over s}\Big({p_{1\mu}p_\perp^2\over\beta-i\epsilon}U+is\partial_\mu U-
{p_{2\mu}\partial_\perp^2U\over\alpha+i\epsilon}\Big){1\over p_\perp^2}U^\dagger
|^{ab}0,[V_i,E^i]^b)
\nonumber\\
&&\hspace{-5mm}
+~{4\over s}\Big({p_{2\mu}p_\perp^2\over \beta-i\epsilon}V+is\partial_\mu V-
{p_{1\mu}\partial_\perp^2V\over\alpha+i\epsilon}\Big){1\over p_\perp^2}V^\dagger
|^{ab}0,[U_i,E^i]^b)
\nonumber
\end{eqnarray}
\subsubsection{ First part of the  Lipatov vertex}

The sum of all the contributions calculated up to now (which includes everything 
but the terms with three-gluon vertex outside the backward cone) can be rewitten 
in the form of Eq. (\ref{vezde})
\begin{eqnarray}
&&\hspace{-5mm}
{\bar A}_i^{{\rm 1st}}~=~\theta(-x_\ast)\theta(-x_\bullet)W_i(x_\perp)
+~ \theta(-x_\ast)\theta(x_\bullet)U_i(x_\perp)
\nonumber\\
&&\hspace{-5mm}
+~\theta(x_\ast)\theta(-x_\bullet)V_i(x_\perp)+
(x|{1\over p^2+i\epsilon p_0}|R^{\rm (1)}_\mu+\delta R^{\rm (2)}_\mu)
\nonumber
\label{Q(1)}
\end{eqnarray}
where $R^{\rm (1)}_\mu(p)$ is given by Eq. (\ref{R(1)})
\begin{eqnarray}
&&\hspace{-0mm}
R^{\rm (1)}_\mu(p)~=~R^{\rm (1)}_{\perp\mu}(p_\perp)+
{2p_{1\mu}\over s}\Big(
{R_{1+}^{\rm (1)}(p_\perp)\over \beta+i\epsilon}
+{R_{1-}^{\rm (1)}(p_\perp)\over \beta-i\epsilon}\Big)
\nonumber\\
&&\hspace{-0mm}+~
{2p_{2\mu}\over s}\Big({R_{2+}^{\rm (1)}(p_\perp)\over \alpha+i\epsilon}
+{R_{2-}^{\rm (1)}(p_\perp)\over \alpha-i\epsilon}\Big)
\label{R(1)aa}
\end{eqnarray}
and the last term 
\begin{eqnarray}
\hspace{-0mm}
\delta R^{\rm (2)}_\mu(k)&=&
{4p_{1\mu}\over s(\beta+i\epsilon)}
([U_i,V^i]V{1\over p_\perp^2}V^\dagger U_j)^{ab}E^{jb}
\nonumber\\
\hspace{-0mm}&-&
{4p_{2\mu}\over s(\alpha+i\epsilon)}
([U_i,V^i]U{1\over p_\perp^2}U^\dagger V_j)^{ab}E^{jb}
\label{deltar}
\end{eqnarray}
is actually a part of the second-order
contribution coming from the  $[U_i,V^i]$ term in the Green function
(see Eq. (\ref{bigo}).

The remaining part of the Lipatov vertex
(coming from the diagram Fig. \ref{fig:2}b with $z$ outside the backward cone) 
will have the same
structure (\ref{R(1)}) with different $R_\perp,R_1$, and $R_2$.
Note that our first part of Lipatov vertex (\ref{R(1)}) satisfies the condition
\begin{equation}
p^\mu R^{\rm (1)}_\mu(p)=p^i R^{\rm (1)}_i+R^{\rm (1)}_{1+}+R^{\rm (1)}_{1-}
+R^{\rm (1)}_{2+}+R^{\rm (1)}_{2-}=0
\label{prov(1)}
\end{equation}
(recall that $(i\partial_i+g[U_i+V_i,)E^i=0$).

\subsubsection{\label{sect:out} The three-gluon vertex outside the backward cone }

Here we must calculate the diagram in Fig. \ref{fig:2}b with 
three-gluon vertex outside the
backward cone $x_\ast,x_\bullet<0$.
With our accuracy, each of the two legs in Fig. \ref{fig:2}b can be represented 
by the field 
\begin{equation}
\bar{A}^{1}_\mu~=~(x|{1\over p^2+i\epsilon p_0}|R^{(1)}_\mu)
\label{Qy}
\end{equation}
so we get
\begin{eqnarray}
\hspace{0mm}
\bar{A}^{(2)}_\mu&=&i\int d^4z(x|(P^2g_{\mu\alpha}+
2iF_{\mu\alpha}+i\epsilon p_0)^{-1}|z)^{ab}
\nonumber\\
\hspace{0mm}
&\times&~
[\bar{A}^{(1)\beta},D_\alpha \bar{A}^{(1)}_\beta-2D_\beta \bar{A}^{(1)}_\alpha]^b(z)
\label{pole1}
\end{eqnarray}
where we have used the gauge condition $D^\mu \bar{A}^{(1)}_\mu=0$ . 

It is easy to see that the term $\sim G_{\mu\alpha}$ in the above equation can be
dropped. Indeed, since the point $z$ lies outside the backward cone, the only
non-vanishing contribution proportional to, say, $F_{\bullet i}$ can come 
from the quadrant $z_\ast\leq 0, z_\bullet\geq 0$ where the only surviving
component of the field is 
$\bar{A}^{(1)}_\ast=-i\delta(x_\bullet)(x_\perp|{1\over p_\perp^2}|[V_i,E^i])$.
(see eq. (\ref{outside})). In addition, 
$D_\bullet\bar{A}^{(1)}_\ast=0$ and therefore all possible terms  
$\sim G_{\mu\alpha}$ in r.h.s. of Eq. (\ref{pole1}) vanish and therefore 
\begin{eqnarray}
&&\hspace{-3mm}
\bar{A}^{(2)}_\mu(x)~=
\label{pole2}\\
&&\hspace{-3mm}i\!\int\! d^4z(x|{1\over P^2+i\epsilon p_0}|z)^{ab}
[\bar{A}^{(1)\nu},D_\mu \bar{A}^{(1)}_\nu-2D_\nu \bar{Q}^{(1)}_\mu]^b(z)
\nonumber
\end{eqnarray}
For the same reason, the Green function ${1\over P^2+i\epsilon p_0}$
in r.h.s. of Eq. (\ref{pole2}) can be replaced by bare propagator
${1\over p^2+i\epsilon p_0}$. 
Indeed, these expressions differ only outside the
forward cone which means either $z_\ast<0,z_\bullet\geq 0$ or 
$z_\ast\geq 0,z_\bullet< 0$ quadrants (recall that we exclude the 
backward cone $z_\ast,z_\bullet< 0$). Consider the contribution
to r.h.s. (\ref{pole2}) coming from $z_\ast<0,z_\bullet\geq 0$ quadrant.
The only nonzero component of the field $A_\mu$ in this quadrant
is $A_\bullet$ (see above) and since $D_\ast A_\bullet=0$ the r.h.s. of
 Eq. (\ref{pole2}) vanishes.
 
We get
\begin{eqnarray}
&&\hspace{0mm}
\bar{A}^{(2)}_\mu(x)~=
\label{pole}\\
&&\hspace{0mm}i\!\int\! d^4z(x|{1\over p^2+i\epsilon p_0}|z)^{ab}
[\bar{A}^{(1)\nu},\partial_\mu \bar{A}^{(1)}_\nu-
2\partial_\nu \bar{A}^{(1)}_\mu]^b(z)
\nonumber
\end{eqnarray}
The Lipatov vertex is represented by the two terms in square brackets. We
will calculate them in turn.

The contribution to Lipatov vertex from the first term is
\begin{eqnarray}
&&\hspace{-5mm}
R^{\rm (2)}_{\mu(1)}(k)~=~
i\!\int\! d^4z~e^{ikz}
[\bar{A}^{(1)\nu},\partial_\mu \bar{A}^{(1)}_\nu](z)
\label{r(2)}
\end{eqnarray}
First, let us calculate the part of this integral coming from the 
product of two $R^{(1)}_\perp$ terms. We have
\begin{eqnarray}
&&\hspace{-4mm}
R^{\rm (2)}_{\mu(11)}(k)~=\int\! {d^4k'\over 16\pi^4}{(k-k')_\mu
[R_\perp^{(1)i}(k'_\perp), R^{(1)}_{\perp i}(k-k'_\perp)]\over 
({k'}^2+i\epsilon k'_0)[(k-k')^2+i\epsilon (k-k')_0]}
\nonumber\\
&&\hspace{-4mm}
=~{i\over 2}\!\int\! {d^2k'_\perp\over 16\pi^3}\!\int_{-\infty}^\infty\!
du~{(1-2u)(\alpha p_1+\beta p_2)_\mu+(k-2k')^\perp_\mu
\over (k'-ku)^2-(k^2+i\epsilon k_0)\bar{u}u}
\nonumber\\
&&\hspace{33mm}
\times~[R_\perp^{(1)i}(k'_\perp), R^{(1)}_{\perp i}(k-k'_\perp)]
\nonumber\\
&&\hspace{-4mm}
=~-{1\over 2}f^{abc}\!\int\! {d^2k'_\perp\over 16\pi^2}{1\over\sqrt{\cal G}}
[R_\perp^{(1)i}(k'_\perp), R^{(1)}_{\perp i}(k-k'_\perp)]
\nonumber\\
&&\hspace{3mm}
\times~\Big\{\Big({p_1\over \beta s}+{p_2\over \alpha s}\Big)(k,k-2k')_\perp
+(k-2k')^\perp\Big\}_\mu
\label{1kycok}
\end{eqnarray}
Note that $k^\mu R^{\rm (2)}_{\mu(11)}(k)~=~0$.

The second part of $R^{\rm (2)}_{\mu(1)}$ is easlily calculated 
using formulas from Appendix D with the result
\begin{eqnarray}
&&\hspace{-3mm}
R^{\rm (2)~a}_{\mu(12)}(k)
\nonumber\\
&&\hspace{-3mm}=
~f^{abc}\!\int\! {d^2k'_\perp\over 16\pi^3}\!\int_{-\infty}^\infty\!
{du\over Z(u)}~
\Big({-2p_{1\mu}\over s(\beta u+i\epsilon)}+~
{2p_{2\mu}\over s(\alpha\bar{u}+i\epsilon)}
\nonumber\\
&&\hspace{-3mm}+~{\bar{u}(\alpha p_1-\beta p_2-k_\perp+2k'_\perp)_\mu\over
(k-k')_\perp^2 (u+i\epsilon)}
\Big)R^{\rm (1)}_{1b}(k'_\perp)R^{\rm (1)}_{2c}(k_\perp-k'_\perp)
\nonumber\\
&&\hspace{-3mm}
+~{4p_{1\mu}\over s(\beta+i\epsilon)}
([U_i,V^i]U{1\over p_\perp^2}U^\dagger)^{ab}[V_j,E^j]^b
\nonumber\\
&&\hspace{-3mm}
-~{4p_{2\mu}\over s(\alpha+i\epsilon)}
([U_i,V^i]V{1\over p_\perp^2}V^\dagger)^{ab}[U_j,E^j]^b
\label{2kycok}
\end{eqnarray}
where $Z(u)=(k'-ku)_\perp^2-(k^2+i\epsilon k_0)\bar{u}u$ and
$R^{\rm (1)}_1\equiv R^{\rm (1)}_{1+}+ R^{\rm (1)}_{1-}$, 
$R^{\rm (1)}_2\equiv R^{\rm (1)}_{2+}+ R^{\rm (1)}_{2-}$ 

The remaining third part of the second-order term is 
\begin{eqnarray}
&&\hspace{0mm}
R^{\rm (2)}_{\mu(1)}(k)~=~
-2i\!\int\! d^4z~e^{ikz}
[\bar{Q}^{(1)\nu},\partial_\nu \bar{Q}^{(1)}_\mu](z)
\nonumber\\
&&\hspace{-3mm}=
~f^{abc}\!\int\! {d^2k'_\perp\over 16\pi^3}\!\int_{-\infty}^\infty\!
{du\over Z(u)}~
\Bigg[-~4{\bar{u}(\alpha p_1-\beta p_2)_\mu\over
(k-k')_\perp^2 (u+i\epsilon)}
\nonumber\\
&&\hspace{-3mm}
\times~
R^{\rm (1)}_{1b}(k'_\perp)R^{\rm (1)}_{2c}(k_\perp-k'_\perp)
\nonumber\\
&&\hspace{-3mm}
+~2(k_i g_{\mu j}-i\leftrightarrow j)
R^{\rm (1)}_{i b}(k'_\perp)R^{\rm (1)}_{j c}
(k_\perp-k'_\perp)\Bigg]
\label{3kycok}
\end{eqnarray}

Combining Eqs. (\ref{1kycok}), (\ref{2kycok}), and
(\ref{3kycok}) 
we get the total second-order term
\begin{eqnarray}
&&\hspace{-3mm}
R^{\rm (2)~a}_{\mu}(k)
\nonumber\\
&&\hspace{-3mm}=
~if^{abc}\!\int\! {d^2k'_\perp\over 16\pi^3}\!\int_{-\infty}^\infty\!
{du\over Z(u)}~
\Bigg[\Big((1-2u)(\alpha p_1+\beta p_2)_\mu
\nonumber\\
&&\hspace{-3mm}
+~(k-2k')^\perp_\mu\Big)R^{\rm (1)~b}_{\perp i}(k'_\perp)
R^{\rm (1)~c~i}_{\perp}(k_\perp-k'_\perp)
\nonumber\\
&&\hspace{-3mm}
+~\Big({-2p_{1\mu}\over s(\beta u+i\epsilon)}+~
{2p_{2\mu}\over s(\alpha\bar{u}+i\epsilon)}
\nonumber\\
&&\hspace{-3mm}+~{\bar{u}(\alpha p_1-\beta p_2-k_\perp+2k'_\perp)_\mu\over
(k-k')_\perp^2 (u+i\epsilon)}
\Big)R^{\rm (1)}_{1b}(k'_\perp)R^{\rm (1)}_{2c}(k_\perp-k'_\perp)
\nonumber\\
&&\hspace{-3mm}
+~2(k_i g_{\mu j}-i\leftrightarrow j)
R^{\rm (1)}_{i b}(k'_\perp)R^{\rm (1)}_{j c}
(k_\perp-k'_\perp)\Bigg]
\nonumber\\
&&\hspace{-3mm}
+~{4\over s}\Big({p_{1\mu}\over \beta+i\epsilon}-
~{p_{2\mu}\over \alpha+i\epsilon}\Big)
\Big([U_i,V^i]U{1\over p_\perp^2}U^\dagger)^{ab}[V_j,E^j]^b
\nonumber\\
&&\hspace{-3mm}
+~([U_i,V^i]V{1\over p_\perp^2}V^\dagger)^{ab}[U_j,E^j]^b\Big)
\label{2orderterm}
\end{eqnarray}
The integral over $u$  yields Eq. (\ref{R(2)}).

 \section{Gluon emission by two Wilson lines
 in the shock-wave background}
\subsection{Classical field induced by a single Wilson line in the shock-wave
background}

In the applications it is sometimes convenient to have the 
result for the classical field and the Lipatov vertex in 
a ``non-symmetric'' form  explicitly expanded 
over the strength of the weak source
$\sim {\rm Tr}\{\rho(x_\perp)U(x_\perp)\}$. This expansion 
corresponds to the diagrams with a gluon production by 
Wilson lines $\parallel p_1$ in the background of the shock wave.
In this case it more convenient to present the results for the covariant 
gauge shock field $A_\bullet\sim \delta(x_\ast)$. (The rotation to the 
pure-gauge field $U_i\theta(-x_\ast)$ is trivial).

In the first order the classical field is given by the two diagrams in 
Fig.\ref{fig:8}
\begin{figure}
\includegraphics{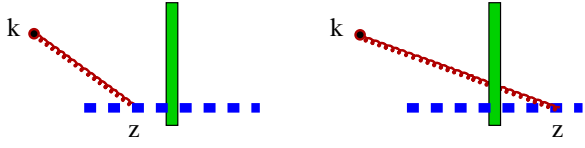}
\caption{\label{fig:8}Emission of a 
gluon by a Wilson line in the shock-wave background.}
\end{figure}
As in Sect. \ref{sect7} we consider here the case of the causal classical field 
corresponding to ${\cal U}=U$ which is the case when we neglect 
the evolution. 
Note it is not difficult to restore the result for  ${\cal U}\neq U$
similarly to Eq. (\ref{fiilds}) - roughly speaking, one should replace 
$(p^2+i\epsilon p_0)^{-1}U$
by  $(p^2+i\epsilon)^{-1}U+2\pi i\delta(p^2)\theta(p_0){\cal U}$.
  
The expression for the classical field produced by one Wilson-line source 
can be read from the (retarded) propagator in a shock-wave background (\ref{retprop1}).
At $x_\ast>0$ one gets
\begin{eqnarray}
&&\hspace{-12mm}
i\langle A^\mu(k)[\infty p_1,-\infty p_1]_z\rangle_{\rm ret}~=~
{1\over k^2+i\epsilon k_0}R^\mu_V(k),
\label{eru1}\\
&&\hspace{-12mm}
R_V^{a\mu}(k)~=~(k|\Big\{2i\partial^\mu U{1\over p_\perp^2}
+ p_1^\mu\Big(2\alpha U{1\over p_\perp^2} 
\nonumber\\
&&\hspace{-5mm}-~{2\over s(\beta+i\epsilon)}U\Big)-{2p_2^\mu\over\alpha s}\partial_\perp^2
U{1\over p_\perp^2}\Big\}|0,z_\perp)^{ab}U_zt^b
\nonumber
\end{eqnarray}
The emission of gluon by the c.c. Wilson line $V^\dagger=[-\infty p_1,\infty p_1]_z$ 
differs from
Eq. (\ref{eru1}) by sign and the replacement $t^b U\leftrightarrow U^\dagger t^b$:
\begin{eqnarray}
&&\hspace{-12mm}
i\langle A^\mu(k)[-\infty p_1,\infty p_1]_z\rangle_{\rm ret}~=~
{1\over k^2+i\epsilon k_0}R^\mu_V(k),
\label{eru2}\\
&&\hspace{-12mm}
R_{V\dagger}^{a\mu}(k)~=~-(k|\Big\{2i\partial^\mu U{1\over p_\perp^2}
+ p_1^\mu\Big(2\alpha U{1\over p_\perp^2} 
\nonumber\\
&&\hspace{-5mm}-~{2\over s(\beta+i\epsilon)}U\Big)-{2p_2^\mu\over\alpha s}\partial_\perp^2
U{1\over p_\perp^2}\Big\}|0,z_\perp)^{ab}t^bU^\dagger_z
\nonumber
\end{eqnarray}

The (transverse) Lipatov vertices in the light-like gauge are obtained from Eq. 
(\ref{lvaxial}):
\begin{eqnarray}
&&\hspace{-5mm}
L^{(1)}_{Vi}(k_\perp;z_\perp)
~=~2(k_\perp|\big[{p_i\over p_\perp^2},U\big]|0,z_\perp)^{ab}U_zt^b
\label{lipu1}\\
&&\hspace{-5mm}
L^{(1)}_{V^\dagger i}(k_\perp;z_\perp)
~=~-2(k_\perp|\big[{p_i\over p_\perp^2},U\big]|0,z_\perp)^{ab}t^bU^\dagger_z
\nonumber
\end{eqnarray}
Note that the fields in this Section are presented 
in the bF gauge in the background of one shock wave
$U$ which differs from the bF gauge for the background field $U_i+V_i$ used in 
the bulk of the paper.
However, the final result (\ref{lipu1}) for the Lipatov vertex $L_i$
corresponds to the $p_2^\mu A_\mu=0$ gauge and therefore agrees with Eq. (\ref{el}). 
Indeed, in Sect. \ref{secteffect} it was
shown that at small $V_i$ Eq. (\ref{el}) reduces to
\begin{eqnarray}
&&\hspace{-2mm}
L^{(1)}_i(k)
~=~2(k|\big[{p_i\over p_\perp^2},U\big]p^k U^\dagger|^{ab}0,V_k^b)
\label{lipsmallv}
\end{eqnarray}
which agrees with Eq. (\ref{lipu1}) if one uses the formula 
$V_xV^\dagger_y=P\exp ig\int_x^ydx^iV_i$ for  
$V(z_\perp)=[\infty p_1+z_\perp,-\infty p_1+z_\perp]$ as in 
Sect. \ref{secteffect}. 

\subsection{Classical field and the Lipatov vertex due to the two Wilson lines}

In the second order, the field due to two Wilson lines is given by the diagrams 
shown in Fig. \ref{fig:9}.
These diagrams are calculated using the retarded Green function (\ref{retprop1}) 
integrated
with the three-gluon vertex. The calculation is similar to that of Appendix C and
the result has the form (the details of the calculation will be published elsewhere):
\begin{eqnarray}
&&\hspace{-2mm}
\langle A^\mu(k)[\infty,-\infty]_z[-\infty,\infty p_1]_{z'}\rangle_{\rm ret}
~=~{R^{(2)\mu}_{VV^\dagger}(k)\over k^2+i\epsilon k_0}
\nonumber\\
&&\hspace{-2mm}R^{(2)a\mu}_{VV^\dagger}(k)~=~
f^{abc}\!\int\!{d^2k'_\perp\over 8\pi^2}
\nonumber\\
&&\hspace{-2mm}
\Bigg\{{1\over\sqrt{\cal G}}\Big(k^\xi\delta_\mu^\eta-k^\eta\delta_\mu^\xi
-{K_\mu\over 2} g^{\xi\eta}\Big)
\tilde{R}_{V_\xi}^b(k'_\perp)\tilde{R}_{V^\dagger\eta}^c(k_\perp-k'_\perp)
\nonumber\\
&&\hspace{-2mm}
+~2i\Big({p_2\over \alpha s}-{p_1\over \beta s}\Big)_{\mu}
\Big[
{(k',k-k')_\perp+{k^2\over 2}-i\sqrt{\cal G}\over {k'}_\perp^2(k-k')_\perp^2} 
\nonumber\\
&&\hspace{-2mm}
\times~
\Big(R_{1V}^b(k'_\perp;z_\perp) R_{2V^\dagger}^c(k_\perp-k'_\perp;z'_\perp)- 
R_{2V}^b(k'_\perp;z_\perp)
\nonumber\\
&&\hspace{-2mm}
\times~ R_{1V^\dagger}^c(k_\perp-k'_\perp;z'_\perp)\Big)
-R^b_{2V}(k';z_\perp)
\nonumber\\
&&\hspace{-2mm}
\times~
r^c_{V^\dagger}(k_\perp-k'_\perp;z'_\perp)+
r^b_V(k';z_\perp) R_{2V^\dagger}(k-k';z')\Big]
\nonumber\\
&&\hspace{-2mm}
-~
2i(g_{\mu\xi}-2{p_{1\mu}k_\xi\over\beta s})
\Big[r^{b\xi}_V(k';z_\perp))r^c_{V^\dagger}(k_\perp-k'_\perp;z'_\perp)
\nonumber\\
&&\hspace{-2mm}-~
r^b_V(k';z_\perp))r^{c\xi}_{V^\dagger}(k_\perp-k'_\perp;z'_\perp)\Big]
\Bigg\}
\label{RV(2)}
\end{eqnarray}
where $K_\mu$ is given by Eq. (\ref{K}), and
\begin{eqnarray}
&&\hspace{-2mm}
R_{1V}^a(k'_\perp;z_\perp)~=~(k'_\perp|[p_\perp^2,U]{1\over p_\perp^2}|z_\perp)^{ab} U_zt^b
\nonumber\\
&&\hspace{-2mm}
 R_{2V}^a(k'_\perp;z_\perp)~=~-
(k'_\perp|\partial_\perp^2U{1\over p_\perp^2}|z_\perp)^{ab} U_zt^b
\nonumber\\
&&\hspace{-2mm}
r_V^a(k'_\perp;z_\perp)~=~(k'_\perp|U{1\over p_\perp^2}|z_\perp)^{ab} U_zt^b
\nonumber\\
&&\hspace{-2mm}
r_V^{b\xi}(k'_\perp;z_\perp)~=~(k'_\perp| i\partial^\xi U{1\over p_\perp^2}|z_\perp)^{bm} U_zt^m
\label{Rrs}
\end{eqnarray}
%
\begin{eqnarray}
&&\hspace{-5mm}
\tilde{R}_V^\xi(k'_\perp;z_\perp)
~=~\Big(2r_V^\xi(k'_\perp;z_\perp)+{(k,k')_\perp+{k^2\over 2}
-i\sqrt{\cal G}\over {k'}_\perp^2}
\nonumber\\
&&\hspace{-5mm}\times~
\Big[{2p_1\over \beta s}R_{V1}(k'_\perp;z_\perp)+
{2p_2\over \alpha s}R_{V2}(k'_\perp;z_\perp)\Big]_\xi\Big)^{ab}
 U_zt^b
\label{eru}
\end{eqnarray}
%
\begin{figure}
\includegraphics{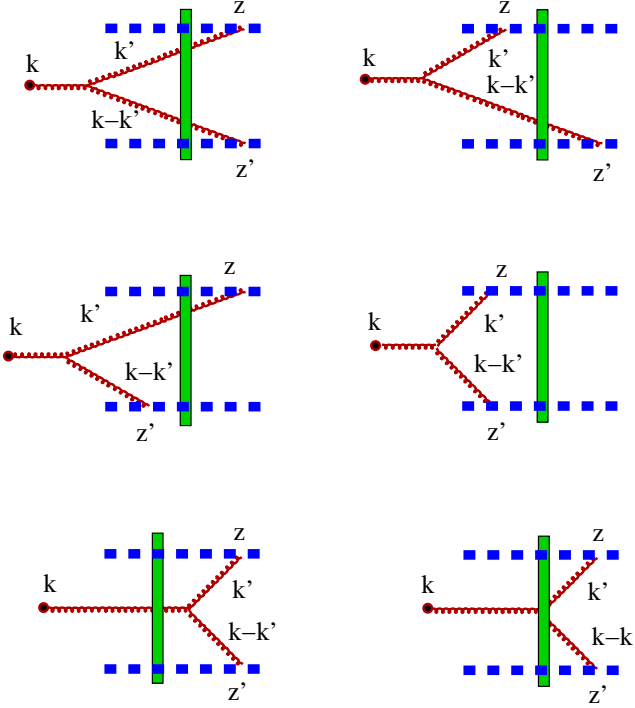}
\caption{\label{fig:9}Gluon emission by two Wilson lines.}
\end{figure}

The corresponding quantities $ R_{V^\dagger}$ and  $r_{V^\dagger}$ are obtained
by substitution $U_zt^b\rightarrow t^bU^\dagger_z$:
\begin{eqnarray}
&&\hspace{-2mm}
R_{1V^\dagger}^a(k'_\perp;z_\perp)
~=~(k'_\perp|[p_\perp^2,U]{1\over p_\perp^2}|z_\perp)^{ab}t^b U_z^\dagger
\nonumber\\
&&\hspace{-2mm}
 R_{2V^\dagger}^a(k'_\perp;z_\perp)~=~-
(k'_\perp|\partial_\perp^2U{1\over p_\perp^2}|z_\perp)^{ab} t^b U_z^\dagger
\nonumber\\
&&\hspace{-2mm}
r_{V^\dagger}^a(k'_\perp;z_\perp)~=~(k'_\perp|U{1\over p_\perp^2}|z_\perp)^{ab} 
t^b U_z^\dagger
\nonumber\\
&&\hspace{-2mm}
r_{V^\dagger}^{b\xi}(k'_\perp;z_\perp)
~=~(k'_\perp| i\partial^\xi U{1\over p_\perp^2}|z_\perp)^{bm} t^mU_z^\dagger
\label{Rdaggers}
\end{eqnarray}
and
\begin{eqnarray}
&&\hspace{-5mm}
\tilde{R}_V^\xi(k'_\perp;z_\perp)
~=~\Big(2r_V^\xi(k'_\perp;z_\perp)+{(k,k')_\perp+{k^2\over 2}
-i\sqrt{\cal G}\over {k'}_\perp^2}
\nonumber\\
&&\hspace{-5mm}\times~
\Big[{2p_1\over \beta s}R_{V1}(k'_\perp;z_\perp)+
{2p_2\over \alpha s}R_{V2}(k'_\perp;z_\perp)\Big]_\xi\Big)^{ab}
 U_zt^b
\label{ervdagger}
\end{eqnarray}
Note that $Ut^a$ and $t^bU^\dagger$ carry the independent indices of 
the Wilson lines 
$[\infty p_1,-\infty p_1]$ and $[-\infty p_1,\infty p_1]$
We do
not display the color indices of $U_zt^b$ and $ t^bU^\dagger_z$ - 
they are always assumed, 
like  $(...)(U_zt^b)^i_j(...)(t^bU^\dagger_z)^k_l$. 
Also, the formula (\ref{RV(2)}) will 
hold true for Wilson lines in the fundamental 
representation provided one replaces 
$(Ut^b)^i_j$ by $(T^bU)^{mn}$ and $(t^bU^\dagger_z)^k_l$ by  
$(T^bU)^{\dagger mn}$.

There is a subtle point in the calculation of diagrams in Fig. \ref{fig:9} 
related to the existence of a term with gluon vertex inside the shock wave.
Consider, for example, the first diagram in Fig. \ref{fig:9}. 
Similarly to Sect. \ref{sect:out}, we calculate the integral over $\beta'$ (the
$p_2$ component of vector $k'$) by taking residues. However, the integral over $\beta'$
becomes divergent if one takes the term $\sim \beta'$ in the three-gluon vertex.
To deal with such divergence, we must retrace one step back and write down the
classical field $\bar{A}^{(2)}$ in the form (\ref{pole1})

\begin{eqnarray}
\bar{A}^{(2)}_\mu&=&i{2p_{2\mu}\over s}\!\int\! d^y(k|{1\over P^2+i\epsilon p_0}|y)^{ab}
\nonumber\\
\hspace{0mm}
&\times&~
[\bar{A}^{(1)\beta},D_\bullet \bar{A}^{(1)}_\beta-2D_\beta \bar{A}^{(1)}_\bullet]^b(z)
\label{prokont1}
\end{eqnarray}
By the equations of motion, one can replace $D_\bullet \bar{A}^{(1)}_\beta$ 
in the r.h.s. of this equation
\begin{equation}
\hspace{0mm}
P_\bullet \bar{A}^{(1)}_\beta ~\rightarrow ~{p_\perp^2\over 2\alpha' s}\bar{A}^{(1)}_\beta+
{2i\over \alpha' s}
\bar{G}_{\bullet \beta}\bar{A}^{(1)}_\ast
\end{equation}
The first term in the r.h.s. of this equation does not produce any 
divergency in $\beta'$ and can be calculated 
by taking residues. 
The second term is a contribution with the point y (position of the three-gluon vertex)
inside the shock wave as shown on the last diagram in Fig. (\ref{fig:9}). Such terms
with the three-gluon vertex inside the shock wave are calculated using the formulas
for the propagator with the initial (or final) points in the shock wave:
\begin{eqnarray}
&&\hspace{-7mm}
(y|{1\over P^2+i\epsilon p_0}|z)~=~[y_\ast,-\infty]_y(y|{1\over p^2+i\epsilon p_0}|z)
\nonumber\\
&&\hspace{-7mm}
(k|{1\over P^2+i\epsilon p_0}|y)~=~
[\infty,y_\ast]_{y_\perp}(k|{1\over p^2+i\epsilon p_0}|y)
\label{inzeshok}
\end{eqnarray}
Summarising, $\beta'$ in the three-gluon vertex must be replaced
by ${{k'}_\perp^2\over\alpha's}$, $\beta-\beta'$ by 
${(k-k')_\perp^2\over(\alpha-\alpha's)}$, and t
the difference must be taken into account
as the term with the gluon vertex inside the shock wave.
It is worth noting that the contribution 
of the last diagram
in Fig. \ref{fig:9} (with the gluon vertex inside the shock wave) is 
essential for the gauge invariance of the Lipatov vertex (cf. ref. \cite{saclay2}).

The classical field due to the two Wilson lines 
$[\infty p_1,-\infty p_1]_z~[\infty p_1,-\infty p_1]_{z'}$ is proportional to 
$R^{(2)}_{VV}(k_\perp;z_\perp,z'_\perp)$
 obtained from (\ref{RV(2)}) by change of sign and the replacement
$R_{V^\dagger}(k-k')^c\rightarrow R_{V^\dagger}(k_\perp-k'_\perp)^c$, 
$r_{V^\dagger}(k_\perp-k'_\perp)^c\rightarrow r_{V^\dagger}(k_\perp-k'_\perp)^c$. 
Similarly,  $R^{(2)}_{V^\dagger V^\dagger}(k_\perp;z_\perp,z'_\perp)$ 
for the classical field due to $[-\infty p_1,\infty p_1]_z~[-\infty p_1,\infty p_1]_{z'}$
is obtained from  (\ref{RV(2)}) by change of sign and the replacement
$R_V^b(k'_\perp)\rightarrow R_{V^\dagger}^b(k'_\perp)$, 
$r_V^b(k'_\perp)\rightarrow r^b_{V^\dagger}(k'_\perp)$.
The Lipatov vertex in the $p_2^\mu A_\mu=0$ gauge takes the form:
\begin{widetext}
\begin{eqnarray}
&&\hspace{-5mm}
\Big(L^{(2)}_{VV^\dagger }(k_\perp;z_\perp,z'_\perp)\Big)^a_i~=~
\lim_{k^2\rightarrow 0}k^2\langle A^a_i(k)
[\infty p_1,-\infty p_1]_z~
 [-\infty p_1,\infty p_1]_{z'}\rangle_{\rm ret}
\label{LV(2)}\\
&&\hspace{-5mm}
=~gf^{abc}
\!\int\!{d^2k'_\perp\over 8\pi^2}\Bigg({1\over\sqrt{k_\perp^2{k'}_\perp^2-(k,k')_\perp^2}}
\Bigg\{\big(k'_i-k_i{(k,k')_\perp\over k_\perp^2}\big)
L^{(1)b}_{Vj}(k'_\perp;z_\perp)
L^{(1)cj}_{V^\dagger}(k-k'_\perp;z'_\perp) 
\nonumber\\
&&\hspace{-5mm} 
+~\Big[
\Big(k_j-k'_j{(k,k')_\perp\over{k'}_\perp^2}\Big)g_{il}
- g_{ij}\Big(k_l-(k-k')_l
{(k,k-k')_\perp\over(k-k')_\perp^2}\Big)\Big]
L^{(1)bj}_V(k'_\perp;z_\perp)
L^{(1)cl}_{V^\dagger i}(k-k'_\perp;z'_\perp) 
\Bigg\} 
\nonumber\\
&&\hspace{-5mm}
+ ~i\Bigg\{\Big(g_{il}{k'_j\over{k'}_\perp^2}-
g_{ij}{(k-k')_l\over(k-k')_\perp^2}\Big)
L^{(1)bj}_V(k'_\perp;z_\perp)
L^{(1)cl}_{V^\dagger i}(k-k'_\perp;z'_\perp)
\nonumber\\
&&\hspace{-5mm}  
-~{2\over {k'}_\perp^2}
\Big(g_{ij}+{k_ik'_j\over k_\perp^2}\Big)R^b_{1V}(k'_\perp;z_\perp)
L^{(1)cj}_{V^\dagger}(k-k'_\perp;z'_\perp)+
{2\over(k-k')_\perp^2}\Big(g_{ij}+{k_i(k-k')_j\over k_\perp^2}\Big)
L^{(1)bj}_V(k'_\perp;z_\perp)
R^c_{1V^\dagger}(k-k'_\perp;z'_\perp)
\nonumber\\
&&\hspace{-5mm}
-~4{k'_ik_\perp^2-k_i(k,k')_\perp\over k_\perp^2{k'}_\perp^2(k-k')_\perp^2}
R^b_{1V}(k'_\perp;z_\perp)R^c_{1V^\dagger}(k-k'_\perp;z'_\perp) 
-2\Big[{k_i\over k_\perp^2}R^b_{2V}(k'_\perp;z_\perp)+
\Big(g_{il}+2{k_ik'_l\over k_\perp^2}\Big)r^{bl}_{1V}(k'_\perp;z_\perp)\Big]
r^c_{V^\dagger}(k-k'_\perp;z'_\perp)   
\nonumber\\
&&\hspace{-5mm}  
+~2r^b_{V}(k'_\perp;z_\perp)
\Big[{k_i\over k_\perp^2}R^c_{2V^\dagger}(k-k'_\perp;z_\perp)
+\Big(g_{il}+2{k_i(k-k')_l\over k_\perp^2}\Big)
r^{cl}_{1V^\dagger}(k-k'_\perp;z_\perp)\Big]
\Bigg\}  \Bigg)
\nonumber
\end{eqnarray}
\end{widetext}
Again, the Lipatov vertex $L^{(2)}_{VV i}(k_\perp;z_\perp,z'_\perp)$
 of the gluon emission by two Wilson lines
$[\infty p_1,-\infty p_1]_z~[\infty p_1,-\infty p_1]_{z'}$
is obtained from (\ref{LV(2)}) by change of sign and replacement 
of $L_{V^\dagger}^c(k_\perp-k'_\perp)$, $R_{V^\dagger}^c(k_\perp-k'_\perp)$ and 
$r_{V^\dagger}^c(k_\perp-k'_\perp)$ by $L_{V}^c(k_\perp-k'_\perp)$
 $R_{V}^c(k_\perp-k'_\perp)^c$ and 
$r_{V}^c(k_\perp-k'_\perp)^c$, respectively. 
The vertex   $L^{(2)}_{V^\dagger V^\dagger i}(k_\perp;z_\perp,z'_\perp)$ 
of the gluon emission by two lines $[-\infty p_1,\infty p_1]_z~[-\infty p_1,\infty p_1]_{z'}$
is obtained from  (\ref{LV(2)}) by change of sign and replacement
of $L_{V}^b(k'_\perp)$, $R_V^b(k'_\perp)$ and $r_V^b(k'_\perp)$ by the corresponding 
vertices $L_{V^\dagger}^b(k'_\perp)$
 $R_{V^\dagger}^b(k'_\perp)$ and $r_{V^\dagger}^b(k_\perp-k'_\perp)$.

\widetext
\mbox{}

\end{document}